\documentclass[apj]{emulateapj}
\def\gtorder{\mathrel{\raise.3ex\hbox{$>$}\mkern-14mu
             \lower0.6ex\hbox{$\sim$}}}
\def\ltorder{\mathrel{\raise.3ex\hbox{$<$}\mkern-14mu
             \lower0.6ex\hbox{$\sim$}}}

\usepackage{colordvi}
\usepackage{amsmath}
\slugcomment{Submitted to PASP}

\shorttitle{LAST overview}

\shortauthors{Ofek et al.}

\begin{document}

\title{The Large Array Survey Telescope -- System Overview and Performances}
\author{E.~O.~Ofek\altaffilmark{1},
S.~Ben-Ami\altaffilmark{1},
D.~Polishook\altaffilmark{2},
E.~Segre\altaffilmark{2},
A.~Blumenzweig\altaffilmark{1},
N.-L.~Strotjohann\altaffilmark{1},
O.~Yaron\altaffilmark{1},
Y.~M.~Shani\altaffilmark{1},
S.~Nachshon\altaffilmark{1},
Y.~Shvartzvald\altaffilmark{1},
O.~Hershko\altaffilmark{2},
M.~Engel\altaffilmark{1},
M.~Segre\altaffilmark{1},
N.~Segev\altaffilmark{1},
E.~Zimmerman\altaffilmark{1},
G.~Nir\altaffilmark{3},
Y.~Judkovsky\altaffilmark{4},
A.~Gal-Yam\altaffilmark{1}, 
B.~Zackay\altaffilmark{1},
E.~Waxman\altaffilmark{1},
D.~Kushnir\altaffilmark{1}, 
P.~Chen\altaffilmark{1},
R.~Azaria\altaffilmark{2},
I.~Manulis\altaffilmark{2},
O.~Diner\altaffilmark{2},
B.~Vandeventer\altaffilmark{5},
A.~Franckowiak\altaffilmark{6},
S.~Weimann\altaffilmark{6},
J.~Borowska\altaffilmark{7},
S.~Garrappa\altaffilmark{6},
A.~Zenin\altaffilmark{6},
V.~Fallah~Ramazani\altaffilmark{6},
R.~Konno\altaffilmark{8},
D.~K\"usters\altaffilmark{6},
I.~Sadeh\altaffilmark{8},
R.~D.~Parsons\altaffilmark{7},
D.~Berge\altaffilmark{7,8},
M.~Kowalski\altaffilmark{7,8},
S.~Ohm\altaffilmark{8},
I.~Arcavi\altaffilmark{8},
R.~Bruch\altaffilmark{1}}

\altaffiltext{1}{Department of particle physics and astrophysics, Weizmann Institute of Science, 76100 Rehovot, Israel.}
\altaffiltext{2}{Department of physics core facilities, Weizmann Institute of Science, 76100 Rehovot, Israel.}
\altaffiltext{3}{University of California, Berkeley, Department of Astronomy, Berkeley, CA 94720}
\altaffiltext{4}{Department of Earth and Planetary Sciences, Weizmann Institute of Science, Rehovot, Israel}
\altaffiltext{5}{Xerxes scientific Ltd.}
\altaffiltext{6}{Faculty of Physics and Astronomy, Astronomical Institute (AIRUB), Ruhr University Bochum, 44780 Bochum, Germany}
\altaffiltext{7}{Institut f\"ur Physik, Humboldt-Universit\"at zu Berlin, D-12489 Berlin, Germany}
\altaffiltext{8}{Deutsches Elektronen-Synchrotron (DESY), D-15738 Zeuthen, Germany}

\begin{abstract}

%        1         2         3         4         5         6         7         8   
%23456789 123456789 123456789 123456789 123456789 123456789 123456789 1234567890

The Large Array Survey Telescope (LAST) 
is a wide-field visible-light telescope array designed to explore the variable and transient sky with a high cadence.
LAST will be composed of 48, 28-cm f/2.2 telescopes (32 already installed) equipped with full-frame backside-illuminated cooled CMOS detectors. 
Each telescope provides a field of view (FoV) of 7.4\,deg$^{2}$ with $1.25$\,arcsec\,pix$^{-1}$, while the system FoV is 355\,deg$^{2}$ in 2.9\,Gpix.
The total collecting area of LAST, with 48 telescopes, is equivalent to a 1.9\,m telescope.
The cost-effectiveness of the system (i.e., probed volume of space per unit time per unit cost) is about an order of magnitude higher than most existing and under-construction sky surveys.
The telescopes are mounted on 12 separate mounts, each carrying four telescopes.
This provides significant flexibility in operating the system.
The first LAST system is under construction in the Israeli Negev Desert,
with 32 telescopes already deployed.
We present the system overview and performances based on the system commissioning data.
The $B_{\rm p}$ 5-$\sigma$ limiting magnitude of a single 28-cm telescope is about 19.6 (21.0), in 20\,s ($20\times20$\,s).
Astrometric two-axes precision (rms) at the bright-end is about 60 (30)\,mas in 20\,s ($20\times20$\,s),
while absolute photometric calibration, relative to GAIA, provides $\sim10$\,millimag accuracy.
Relative photometric precision, in a single 20\,s (320\,s) image,
at the bright-end measured over a time scale of about 60\,min
is about 3 (1)\,millimag.
We discuss the system science goals, data pipelines,
and the observatory control system in companion publications.

\end{abstract}


\section{Introduction}
\label{sec:Introduction}

Scanning the sky repeatedly has revealed
intriguing facts about the physical Universe.
From the detection of the Earth's precession
by Hipparchus, to the discovery of the proper motion of stars by Halley, and the extragalactic novae of Fritz Zwicky.
With the advance of technology, sky surveys have made great progress in the past 20 years.
Sky surveys revealed new kinds of exploding phenomena
like super-luminous supernovae (\citealt{Quimby+2007_SN2005ap_SLSN}; \citealt{Quimby+2011_SLSN}),
discovered thousands of exoplanets around distant stars (e.g., \citealt{Zhu+Dong2021ARA&A_Exoplanets_Statistics_Implications}), 
measured the distances to a large number of stars (e.g., \citealt{GAIA+2016_GAIA_mission}),
and
identified the majority of Near Earth Objects larger than 1\,km.

While sky-surveys like the Zwicky Transient Facility (ZTF; \citealt{Bellm+2019_ZTF_Overview}), Pan-STARRS (\citealt{Chambers+2016_PS1_Surveys}), ASAS-SN (\citealt{Kochanek+2017_ASASSN_VarStars}), and ATLAS
(\citealt{Heinze+2018_ATLAS_VarStars})
provide excellent monitoring of the sky, making progress in our understanding of the Universe requires pushing towards observing a larger fraction of the sky continuously (i.e., around the globe) at high cadences.
Furthermore, it is desirable to increase the cost-effectiveness of survey telescopes, otherwise, in the future, it will be expensive and difficult to surpass the performances of the Large Survey of Space and Time (LSST; \citealt{Abell+2009_LSST_ScienceBook_Ver2}, \citealt{Ivezic+2019_LSST_Survey}).
Here, following \cite{Ofek+BenAmi2020_Grasp_SkySurvrys_CostEffectivness},
we define the cost-effectiveness of a survey telescope as the relative volume of the Universe that can be observed by a system per unit time per unit cost (i.e., grasp per unit cost).
\cite{Ofek+BenAmi2020_Grasp_SkySurvrys_CostEffectivness}
argued that
with the availability of affordable back-side-illuminated CMOS detectors with small pixels (smaller than about 4\,$\mu$m),
and taking advantage of the affordability of off-the-shelf components,
it only now became significantly more cost-effective to construct survey telescopes composed of multiple small telescopes,
compared with a single large telescope with the same grasp (relative volume per unit of time).
In fact, the increase of small telescope cost-effectiveness is so pronounced, that we believe
that, with current technology, there is no point anymore in building telescopes larger than about 0.5-m for seeing-limited visible-light imaging purposes.
Another future project that has the potential to demonstrate this point is
the Argus array (\citealt{Law+2022PASP_ArgusArray}).

Following these lines of argument, we are constructing the first system (also referred to as a node) of 
a new cost-effective survey telescope: the Large Array Survey Telescope\footnote{\url{https://www.weizmann.ac.il/wao/}} (LAST).
LAST is currently the sky survey system with the highest grasp
(Figure~\ref{fig:Grasp}).
In the near future, both LSST (\citealt{Ivezic+2019_LSST_Survey}) and ULTRASAT (\citealt{Sagiv+2014_ULTRASAT}, Shvartzvald et al., in prep)
will surpass these grasp.
Nevertheless, LAST geographic position
elevate our capabilities in terms of monitoring the sky
around the globe.
We estimate that LAST
cost-effectiveness is about an order of magnitude higher compared to most other survey telescopes.

Here we describe the LAST system, design, strategy, and preliminary performances.
In companion papers, we discuss the LAST science goals (\citealt{Ben-Ami+2023PASP_LAST_Science}),
pipeline (\citealt{Ofek+2023ASP_LAST_PipelineI}; hereafter pipeline paper), and observatory control (Segre et al., in prep.).
The first science results from LAST, including analysis of the DART (\citealt{Rivkin+2021PSJ_DART_MissionRequirments}) impact observations, are described in Ofek et al., submitted.

In \S\ref{sec:Description} we provide an overview of the system, while in \S\ref{sec:DataRate}
we discuss data rate considerations
that dictate the survey and pipeline strategy.
In \S\ref{sec:Strategy} we discuss the survey strategy,
and in \S\ref{sec:Software} we provide an overview of the system software.
In \S\ref{sec:Cam} we describe the measured camera
properties, while in \S\ref{sec:Perform} we present
some of the system's initial performances.
Future follow-up facilities to support LAST are discussed
in \S\ref{sec:FollowUp},
and we conclude in \S\ref{sec:Conclusions}.

\section{LAST system overall description}
\label{sec:Description}

LAST
is designed to be a cost-effective, modular, and extendable survey telescope.
A full LAST node is composed of 48 ($=N_{\rm tel}$), 28-cm
telescopes under a single rolling-roof structure.
Each telescope (\S\ref{subsec:Telescopes}) is equipped with
a full frame (i.e., $36\times24$\,mm-size)
backside illuminated and cooled CMOS detector
(\S\ref{subsec:Cameras}).
A group of four telescopes are mounted on a single, two-sided,
German Equatorial mount (see \S\ref{subsec:Mounts}).
%In addition, each mount
%is equipped with auxiliary sensors (e.g., temperature, humidity, wetness, vibrations, accelerometers; see \S\ref{subsec:Sensors}).
Each sub-system of four telescopes is controlled
by two 30-core 256GB RAM
computers
(see \S\ref{subsec:Computers}).
In addition, these computers are responsible for
running the image processing pipeline (\S\ref{subsec:pipeline}; \citealt{Ofek+2023ASP_LAST_PipelineI}).

\begin{figure}
\centerline{\includegraphics[width=8cm]{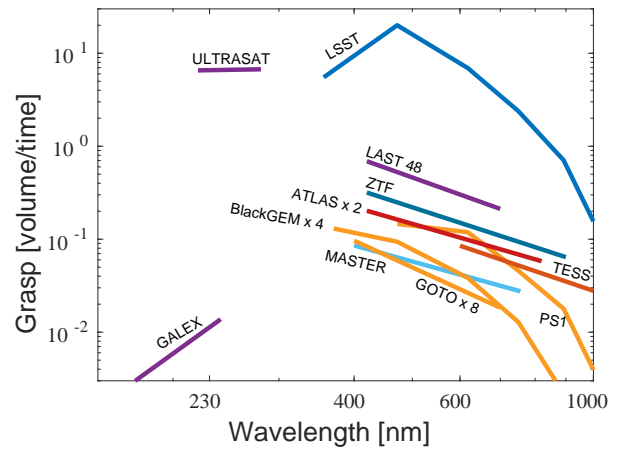}}
\caption{The grasp (relative volume per unit time) of several sky surveys, as a function of wavelength,
as calculated using the used or planned survey exposure time, and published limiting magnitudes
(rather than aperture, seeing, and sky brightness).
The plot is calculated for a source with 20,000\,K black-body
spectrum (e.g., hot transient).
The plot also takes into account the on-sky fraction of time (about $0.25$ for a typical ground-based observatory),
and the survey exposure time and dead time (if available).
The y-axis is in arbitrary units, normalized such that the LAST grasp is about 1.
LAST 48, assumes a LAST node with 48 telescopes.
Black-GEM $\times4$, assumes four Black-GEM telescopes (\citealt{Bloemen+Groot+2015ASPC_BlackGEM}),
GOTO $\times8$ assumes the GOTO system with eight telescopes (\citealt{Steeghs+2022MNRAS_GOTO_TelescopeSurvey}),
for ATLAS we assume two telescopes (\citealt{Tonry2011_ATLAS_SurveyCapability}),
while MASTER is calculated assuming seven sites.
The Argus array-like project (\citealt{Law+2022PASP_ArgusArray})
has the potential to be at the level of LSST to about an order of magnitude above LSST.
However, we do~not know
the exact expected parameters for this system.
\label{fig:Grasp}}
\end{figure}

A summary of a LAST node system parameters
are provided in Table~\ref{tab:system}.
\begin{deluxetable}{ll}
\tablecolumns{2}
\tablewidth{0pt}
%\tabletypesize{\footnotesize}
\tablecaption{The first LAST node system's parameters}
\tablehead{
\colhead{Property}    &
\colhead{Value}      \\
\colhead{}       &
\colhead{}  
}
\startdata
Number of telescopes (planned; Jun 2023)     & 48  \\
Number of telescopes (Mar 2023)     & 32  \\
Telescopes per mount      & 4 \\
Telescope aperture        & 279.4\,mm \\
System equivalent aperture & 1.9\,m ($\cong 0.28\sqrt{48}$)\\
Telescope focal length    & 620\,mm \\
Pixel scale               & $1.25''$\,pix$^{-1}$ \\
Telescope FoV             & $2.2\times 3.3$\,deg$\cong7.4$\,deg$^{2}$  \\
System FoV                & 355\,deg$^{2}$ \\
Total number of pixels    & $\cong 2.9\times10^{9}$\\
$B_{\rm p}$ Limiting magnitude ($5\,\sigma$ in 20\,s) & $\approx19.6$\,mag \\
$B_{\rm p}$ Limiting magnitude ($5\,\sigma$ in $20\times20$\,s) & $\approx21.0$\,mag\\
Location         & Neot-Smadar, Israel \\
Longitude (WGS84)& 35.0407331\,deg E \\
Latitude (WGS84) & 30.0529838\,deg N \\
Height (WGS84)   & 415\,m
\enddata
\tablecomments{Limiting magnitude is estimated during dark time, and airmass of about 1, image quality of $\approx2.8''$, for sources with color of $B_{\rm p}-R_{\rm p}=1.0$\,mag near the field center (see \S\ref{subsec:LimMag} for details).}
\label{tab:system}
\end{deluxetable}
In Figure~\ref{fig:Grasp} we present the estimated grasp,
as a function of wavelength, for several
survey telescopes for which the system parameters are known to us.
The total hardware and construction costs for
a LAST node is \$1.4M.
This includes all the telescopes, cameras, mounts, computers, communication, and construction
costs including the enclosure, and site infrastructure.
The man-power cost for the system development
(software and hardware) is estimated at about \$500k.
The LAST PIs are Eran Ofek \& Sagi Ben-Ami.

\begin{figure}
\centerline{\includegraphics[width=8cm]{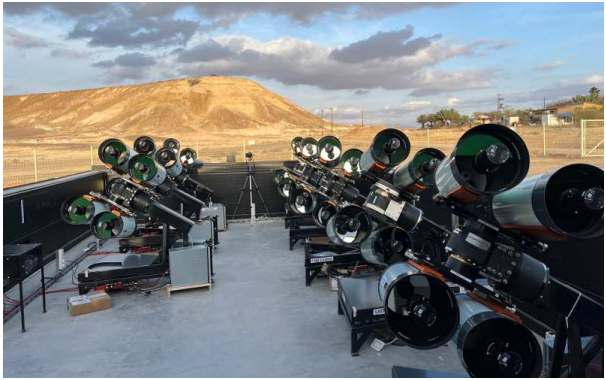}}
\caption{LAST with 32 telescopes installed. The enclosure is fully open.
\label{fig:LAST_12telescopes}}
\end{figure}

The LAST system is highly modular,
both in terms of hardware and software.
This modularity as well as the relatively small physical size of the components makes the system flexible,
but it also makes it easy to extend, deploy,
and maintain.
Almost all the components (with the exception
of the enclosure and mounts) are off-the-shelf products that are manufactured in large numbers.
Figure~\ref{fig:LAST_12telescopes} shows an image of the first LAST node with 12 telescopes,
in Neot-Smadar.

\subsection{Site}
\label{subsec:Site}

The sky brightness in the Neot-Smadar site is relatively poor.
In the past two years the $V$-band sky brightness degraded from about 21.0\,mag\,arcsec$^{-2}$
to about 20.6\,mag\,arcsec$^{-2}$.
A major consideration for the site selection was the speed at which the observatory can be built in the site (e.g., permits).
We note that building LAST in a dark site on Earth (i.e., 22\,mag\,arcsec$^{-2}$),
may improve the system limiting magnitude by about 0.7\,mag.

\begin{figure}
\centerline{\includegraphics[width=8cm]{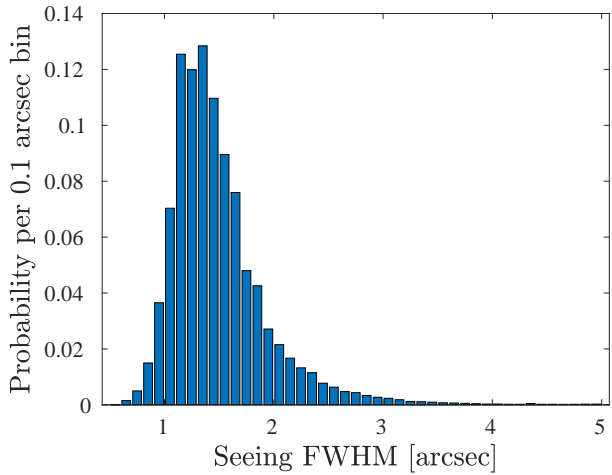}}
\caption{Histogram of about 75,000 seeing measurements taken over 180 nights prior to the observatory construction.
The mode, median, and mean seeing measured over $180$ nights is about 1.3, 1.4, and 1.5\,arcsec, respectively.
The lower and upper 5\% percentile of the measurements is 0.98 and 2.36\,arcsec, respectively.
\label{fig:NeotSmadarSeeing}}
\end{figure}

Although LAST is marginally seeing-limited,
we are planning to use this site for additional
telescopes (\S\ref{sec:FollowUp}), and therefore we have measured the site seeing extensively.
The seeing was mainly measured using a Cyclope\footnote{https://www.alcor-system.com/new/SeeingMon/Cyclope.html} seeing monitor device. The site mode, median, and mean seeing measured over $180$ nights is about 1.3, 1.4, and 1.5\,arcsec, respectively.
The histogram of all seeing measurements is shown in Figure~\ref{fig:NeotSmadarSeeing}.

\subsection{Enclosure}
\label{subsec:Enclosure}

The LAST enclosure is a custom-made rolling-roof structure.
The inner usable size of the structure is 5.2\,m by 11\,m. 
The roof is composed of three parts that move in the direction of the long axis of the structure.
When the roof is open, the structure walls are 1.2\,m high, with the exception of the enclosure entrance.
The structure has additional upward/downward-moving walls on three sides (South, East and North). These walls can go up in order to block strong winds.
The LAST enclosure was designed to be an affordable structure.
The construction of the enclosure on site took about six days.

\subsection{Telescopes}
\label{subsec:Telescopes}

The LAST telescopes are Celestron \textregistered 11-inch
f/2.2 (focal length 620mm) Rowe-Ackermann-Schmidt
telescopes.
The telescopes' optical elements include an aspheric corrector,
spherical mirror and a four-elements field flattener, and are designed to produce image quality of about $1.6''$ over the field of view of a full-frame sized camera.

Similar telescopes are also available in an
8-inch\footnote{The 8-inch version, uses a different optical design with f/2 and its image quality is worse than the 11 and 14-inch telescopes.}
and 14-inch version.
In terms of grasp or etendue cost-effectiveness, and given the image quality delivered by these telescopes,
the 11-inch version provides the best performance.
Even if we ignore the price aspect,
assuming a constant-size detector,
the grasp of a single 11-inch telescope 
is about 35\% larger than that of the 14-inch telescope.

\subsection{Mounts}
\label{subsec:Mounts}

We use the Xerxes equatorial mount. This mount uses two direct-drive Kollmorgen \textregistered~motors that can produce torques of 60\,N-m, and 33\,N-m, for the hour angle and declination, respectively. The motors are equipped with
Renishaw \textregistered~encoders and are
controlled by two Copley \textregistered~controllers.
The controllers are commanded via a serial link with
an interface developed by our group.
The mounts can move at a very high speed, but for safety reasons, we limit their
operations to speeds of up to 12\,deg\,s$^{-1}$.

\subsection{Cameras}
\label{subsec:Cameras}

In the focal plane of each telescope,
we use the QHY600-PH camera,
hosting a cooled SONY\textregistered~IMX455 backside-illuminated CMOS detector.
The camera and detector properties are summarized in Table~\ref{tab:camera} and Table~\ref{tab:cameraPars}.
In terms of Gain and Readout noise, the detector has several modes,
and it is possible to control its gain
and offset (bias level).
We choose to work with the 16bit mode,
with a specific gain/offsets,
which are listed in Table~\ref{tab:cameraPars} (see \S\ref{subsec:Gain}).

The camera uses a rolling shutter and hence can be read continuously, with negligible dead time between exposures (video mode).
In the video mode, it takes about 0.7\,s
to read the image from the camera to the computer memory, and an additional $\sim1.5$\,s to write the image to a spinning hard drive.
Therefore, in continuous observing mode, exposures as short as 0.8\,s can be used.

The LAST cameras can be equipped with a single
non-exchangeable filter.
Indeed we are planning to test the use of such filters
and polarimetry filters in the future.
However, currently, we do~not equip the cameras with any filters\footnote{Some tests are being performed with polarization filters.}.
The reasoning for this is that LAST is designed to be a discovery machine, while followup (multi-band imaging and spectroscopy) will be conducted using other telescopes (see \S\ref{sec:FollowUp}).

\begin{deluxetable}{ll}
\tablecolumns{2}
\tablewidth{0pt}
%\tabletypesize{\footnotesize}
\tablecaption{QHY600M-PH camera}
\tablehead{
\colhead{Property}    &
\colhead{Value}      \\
\colhead{}       &
\colhead{}  
}
\startdata
Model      & QHY600M-PH  \\
Sensor     & IMX455 \\
Size       & $36\times24$\,mm \\
Pixel size & $3.67\mu$m \\
Pixels     & $60.8\times10^{6}$   \\
Sensitive pixels & $6354 \times 9576$ \\
Cooling    & Two stage thermoelectric\\
Readout time (buffer to memory) &  $\approx0.7$\,s \\
Readout mode & Rolling shutter
%\hline
%Mode/Gain/offset setting & 2 / 0 / 4 \\
%Gain          &   0.8\,e$^{-}$/ADU \\
%Mean Readout noise & 4.5\,e$^{-}$  \\
%Dark current at $-5$C &  $\sim8\times10^{-3}$\,e$^{-}$\,pix$^{-1}$\,s$^{-1}$ \\
%Bias level    &  74 ADU \\
%Full well     &  $\approx 5\times10^{4}$\,e$^{-}$
\enddata
\tablecomments{The QHY600M-PH camera properties. Additional parameters are listed in Table~\ref{tab:cameraPars}}
\label{tab:camera}
\end{deluxetable}

With the LAST telescope,
the camera provides a plate scale of $1.25''$/pix,
and a field of view of $3.3\times2.2$\,deg
($\cong7.38$\,deg$^{2}$).
By default the telescopes and cameras, on each mount,
are set to observe a contiguous $\approx 6.5\times4.3$\,deg
field of view.
The cameras are set to have about 10\,arcmin overlap,
with about 3\,arcmin accuracy.
A set of shims allow us to switch between the open mode (i.e., wide field default),
and the narrow mode (all four telescopes are pointing to the same direction).
All the cameras are aligned such that their long-axis is oriented North-South to an accuracy of about 1\,deg.

The cameras are controlled using the QHY software development kit (SDK), using custom software developed by our group.
This software, as well as the observatory control system, is described in Segre et al., in prep.

\subsection{Focusers}
\label{subsec:Focusers}

Each telescope is equipped with a Celestron electric focuser. The focusers are connected to the operating computer via USB links.
However, to increase reliability, a new focuser system is being designed.
The telescopes are focused by minimizing
the width of the point spread function (PSF)
of a large number of high $S/N$ stars near the center of the field.
This is done, once, in the beginning of each night (see \S\ref{subsec:FocusTipTilt}).
However, since the focus is temperature dependent,
we need to refocus the telescope during the night.
This refocusing is done using a simple temperature-focus model, and this is activated, if the mount temperature varies by more than 1 degrees Celsius, averaged over ten minutes, since the last refocus.

\subsection{Computers}
\label{subsec:Computers}

A LAST node includes two computers
per mount, plus a single control/manager computer per node.
The two computers per mount are responsible for 
image processing (\S\ref{subsec:pipeline}),
controlling the mount (one of the computers),
and controlling the cameras and focusers (two cameras/focusers per computer).
All the computers are running Linux Ubuntu.
Since all the computers should be identical in setup and software installation, we have developed a custom software installation tool ({\tt last-tool}) written in {\tt bash}.
The tool is responsible for checking and enforcing the software policy on all computers.

Each computer has 30 cores with 256\,GB RAM.
In addition, each computer has a 256\,GB SSD disk,
6\,TB disk for software and {\tt catsHTM} catalogs (\citealt{Soumagnac+Ofek2018_catsHTM}),
and two removable 14\,TB disks for imaging data.
Most of the CPUs are used by the image analysis pipeline.

The observatory manager computer
is responsible for running the scheduler,
and allocating targets to the individual mounts.
This computer also controls the dome, weather stations,
and auxiliary equipment.
The computers are connected via a local network ($1-10$\,Gbit\,s$^{-1}$), that is by itself connected to the internet using a 500\,Mbit\,s$^{-1}$ fiber link.

\subsection{Auxiliary sensors}
\label{subsec:Sensors}

LAST is equipped with several auxiliary sensors.
First, each mount has its own
temperature sensor.
The temperature sensor is located on the mount, near the telescopes, and its primary goal is to provide temperature readings for the focuser system
(see \S\ref{subsec:Focusers}).

Next, the main observatory control computer is connected to several meteorological and all-sky camera sensors,
that are used by the observatory control system to determine when to it is safe to observe.

\section{Data rate considerations}
\label{sec:DataRate}

A major consideration for the LAST observing strategy and pipeline design is related to the system data rate.
Each LAST camera produces a 16-bit 61\,Mpix image.
Due to the requirements of
several science goals
(e.g., exoplanets around white dwarfs; see \citealt{Ben-Ami+2023PASP_LAST_Science}), we choose a nominal exposure time of 20\,s.
This is roughly half the expected duration of an exoplanet transiting a white dwarf. 
Furthermore, in our system and site, the transition\footnote{We define this transition when the background variance is equal to the read-noise squared.}
from 
readnoise dominated noise to background-dominated noise takes place at exposure times of about 5\,s.
With our nominal exposure time, we are in the background-dominated noise regime.
Therefore, image coaddition does~not suffer from significant losses.
For example, by combining 20 images we gain 1.4\,mag instead of 1.6\,mag (i.e., $\cong2.5\log_{10}{\sqrt{20}}$) improvement in limiting magnitude.

This exposure time results in a data rate of
65\,Mbit\,s$^{-1}$ per camera and 2.2\,Gbit\,s$^{-1}$ for the entire array during operations, taking into account the telescopes-mount slew time and restarting the video mode
in the beginning of each visit.
This data rate is relatively high\footnote{This data rate is roughly 70\% higher than the Vera Rubin telescope (\citealt{Ivezic+2019_LSST_Survey}) data rate (assuming 30\,s exposures),
about ten times higher than the ZTF data rate (\citealt{Bellm+2019_ZTF_Overview}),
and about $1/3$ of the W-FAST (\citealt{Nir+Ofek+2021_WFAST}) data rate.}.
Such data rate is expected to generate about
$2$\,PB\,yr$^{-1}$ of raw images.
Given the costs associated with storing and transferring such a large amount of data, we choose a strategy that will allow us, on one hand, to reduce (if needed) the amount of stored data, and on the other hand to keep the high cadence temporal information.
This high data rate also means that it is desirable to perform most of the data processing on-site.

Our default strategy is to analyze the data from
each telescope separately, and let each telescope
observe each field for $20\times20$\,s ($=400$\,s).
Among the advantages of our 20-images per field
strategy are:
the ability to screen satellite glints
in a single visit (\citealt{Corbett+2020_SatellitesGlints}; \citealt{Nir+2020_Satellites_Glints_FlaresLimit}; \citealt{Nir+2021_RNASS_GN-z11-Flash_SatelliteGlint}),
and to detect main belt asteroids using their motion
in about 6\,min (see \citealt{Ofek+2023ASP_LAST_PipelineI}).
Furthermore, for short-time-scale phenomena like exoplanets around white dwarfs and flaring stars, a single visit provides a light curve of these events.

In the long-term archive, we keep only
the individual image catalogs, and the coadd images
(as well as other data products based on the coadded image, e.g., catalogs and subtraction images).
However, the individual raw images
are stored locally (on-site) in a cyclic buffer
for a period of about 60\,days.

This strategy allows us to reduce the data rate
by a factor of about 8, but keep some of the high cadence information.
In addition, if needed, the individual images can be saved from deletion.
In the future we may save also the individual raw images.

\section{Observing strategy}
\label{sec:Strategy}

The multi-mount, multi-telescope structure
of LAST offers great flexibility in terms of operations.
One question is whether to use
the wide mode (28-cm telescope with 355\,deg$^{2}$),
or a narrow mode (e.g., 1.9\,m telescope with 7.4\,deg$^{2}$).
In terms of grasp, operating
the LAST system in the wide mode provides a grasp that is about 2.6 ($=N_{\rm tel}^{1/4}$) times larger compared with the grasp of the system operated at the narrow mode.
Therefore, our primary strategy is to use the wide mode,
and to set the default pointing of four telescopes
on each mount to cover a wide but contiguous field of view.
However, for some specific science cases,
the narrow mode has advantages (e.g., precision photometry; see \citealt{Ben-Ami+2023PASP_LAST_Science}).

The LAST observing strategy is flexible and likely to change based on specific science goals,
and capabilities.
Our observing strategy can be roughly divided
between three programs:
(i) high cadence; (ii) low cadence;
and (iii) target of opportunity (ToO).
In the following discussion, we are going to assume that the ToO program will
take about 5\% of the observing time and that on average (all year) we have 6 hours of clear sky every night.
Assuming a visit of $20\times20$\,s our scanning speed is about 1470\,deg$^{2}$
per mount (four telescopes) per night (6 hours).
This number is 17,640\,deg$^{2}$ per array (12 mounts).
Assuming a visit of $5\times20$\,s  our scanning speed is about 4900\,deg$^{2}$
per mount per night,
or 58,800\,deg$^{2}$ per array.
The latter visit duration, therefore, allows us to scan about 9800\,deg$^{2}$\,hr$^{-1}$
down to a limiting magnitude of about 20.3 on each visit.

For several reasons including science goals and pipeline performance,
in the first six months of the project, we are likely to implement
the $20\times20$\,s visits, with a slow$+$fast cadence survey strategy.
In this strategy, we will use $1/3$ of the telescopes to observe
about 5900\,deg$^{2}$ twice a night every two nights
and about 1470\,deg$^{2}$ eight time per night, every night.
Later on, we will consider using the $5\times20$\,s visits,
and cover about $10^{4}$\,deg$^{2}$ every hour.
The ToO time will be used to respond to GW, neutrinos, and GRB triggers (e.g., \citealt{Abbott+2017_GW170817_MultiMessengerObservations}, \citealt{Icecube+2017A&A_Icecube_NeutrinoMultipletsSearch}).
ToOs interrupt the main program for all or some of the mounts as needed.

\section{Software}
\label{sec:Software}

The LAST project required major software efforts.
These include, the pipeline (\S\ref{subsec:pipeline});
the observatory control system (\S\ref{subsec:ocs});
the scheduler (\S\ref{subsec:scheduler});
and apparatus calibration software (\S\ref{subsec:calib}).
Additional software tools include {\tt last-tool} -- a software package to enforce the LAST software installation policy on all the LAST computers.

\subsection{Pipeline}
\label{subsec:pipeline}

The LAST data reduction pipeline is described in \cite{Ofek+2023ASP_LAST_PipelineI}.
Here we provide a brief overview of the pipeline.
The on-site operations
include: a basic calibration
(e.g., dark subtraction, flat correction,
bit-mask production),
background estimation, source detection,
astrometry, photometric calibration,
matching sources in all the exposures from a visit,
searching for variable sources (flares/transits),
and searching for moving sources.
The visit exposures are then registered and coadded.
Next, for each coadded image, we estimate the background and variance,
propagate the mask images, find and measure sources including PSF photometry,
refine the astrometry,
and match the sources against external catalogs.
In addition, we perform a quick (catalog-based) transient detection.
Next, the coadd images are transferred to the Weizmann Institute campus,
via the internet, and image subtraction and transient detection
is performed
using the \cite{Zackay+2016_ZOGY_ImageSubtraction} algorithm.
Reference images are built using the proper coaddition algorithm (\citealt{Zackay+2017_CoadditionI}; \citealt{Zackay+2017_CoadditionII}).

In order to reduce costs and complexity it was critical to reduce the amount of computing (and electricity) power on site.
Using off-the-shelf software packages will require an order of magnitude increase in the amount of computing power on site.
Therefore, we have developed a new efficient image processing code for LAST and ULTRASAT (\citealt{Sagiv+2014_ULTRASAT}, Shvartzvald et al., in prep.).
For example, our source extraction code which performs source finding
and PSF photometry is about a factor of 30 faster than
{\tt SExtractor} (\citealt{Bertin+1996_SExtractor}).
The code is based on the tools developed in \cite{Ofek2014_MAAT},
\cite{Soumagnac+Ofek2018_catsHTM}, \cite{Ofek2019_Astrometry_Code},
and is available via GitHub\footnote{https://github.com/EranOfek/AstroPack}.

\subsection{Observatory Control System}
\label{subsec:ocs}

LAST is operated by the LAST Observatory Control System (OCS), which is described in detail in Segre et al., in prep.
The control software has two main levels -- a Unit Control System (UCS) that is responsible for operating a single mount and its four telescopes and cameras,
and an , under construction, Observatory Control System (OCS)
that is responsible for the health of the observatory and allocating tasks to the 12 UCS.
Additional tasks of the OCS include the control of the enclosure, and making the decisions to open and close the observatory based on the available safety, weather, and security information.

\subsection{Targets selection}
\label{subsec:scheduler}

\subsubsection{The Scheduler}
\label{subsubsec:sched}

Each LAST mount is designed to be operated in two ways:
(i) as an independent, stand-alone mount,
which gets its targets from a list or a scheduler;
(ii) part of a collective of mounts,
that get their targets from a scheduler.
The main difference between the two
ways of operation is that central scheduling requires also a function
that allocates targets to each mount.
The allocation process is not straight-forward because
different mounts have different sky visibility constraints (see below).

The LAST scheduler may have three main sources of targets:
a predefined list of fields to observe,
Target of Opportunity (ToO) and user-inserted targets.
The LAST scheduler is designed to satisfy several criteria:
(1) To observe the targets according to the requested global cadence, and nightly cadence (i.e., number of visits during the night);
(2) To schedule the observations of a target such that the airmass during the observations is minimized;
and (3) To avoid fields near the Moon.
The target assignment to different mounts is performed by the allocator (\S\ref{subsubsec:alloc}).

\begin{figure}
\centerline{\includegraphics[width=7.5cm]{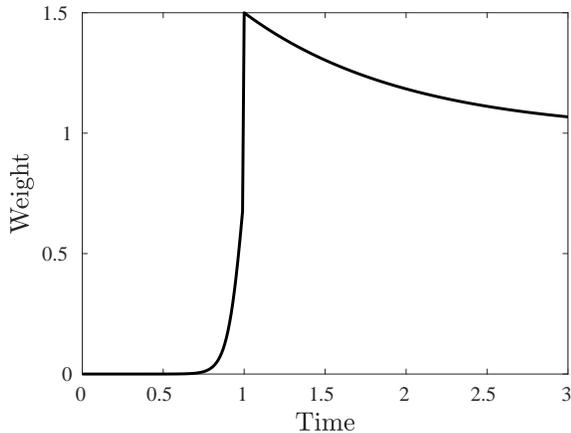}}
\caption{A schematic plot of our default weight function,
as a function of the time since the last visit taken on previous nights.
The rising part of this function is modeled as a Fermi-Dirac function,
while the decaying part is an exponential decaying asymptotically to 1.
\label{fig:WeightTime}}
\end{figure}

The global and nightly cadence are achieved by using a specific weight function, that is calculated
as a function of: the time from the last previous-night observation of the target;
the time of the latest observation on the same night; and the number of observations during the night
that were already executed.
Figure~\ref{fig:WeightTime} shows a schematic plot of our default weight as a function of the time since the last visit taken on previous nights.
The rising parts of this function are modeled as a Fermi-Dirac function,
while the decaying part is an exponential decaying asymptotically to 1.
The reasoning for the decaying part
is that we prefer to observe some fields with a regular cadence, rather than all fields with a poor cadence.

On top of that, the weight of each target is multiplied by a binary visibility
for each target.
The visibility is calculated in the following way:
(i) If the target has an angular distance to the Moon which is smaller than a Lunar-illumination
dependent threshold, then the visibility of the target is set to zero;
(ii) A target is regarded as visible if it is observable for at least 2 (5) hours during the night with an airmass larger than 2,
for the slow (high) cadence.
In addition, the visibility time window is chosen to minimize the target airmass.

Finally, the scheduler deals with the fast-cadence and low-cadence fields separately.
The operator decides how many mounts are allocated for the fast cadenced program,
and how many mounts for the slow cadence program.
This method simplifies the scheduling of different programs.

\begin{figure}[h]
\centerline{\includegraphics[width=7.5cm]{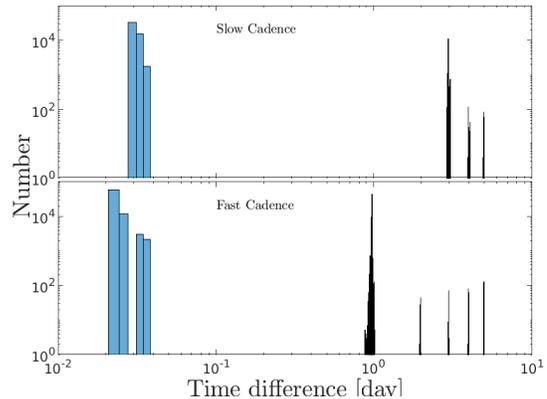}}
\caption{Histogram of one year simulated time difference between visits in the slow-cadence survey (upper panel) and fast-cadence survey (lower panel). The lunar cycle and length of nights are taken into account, but the weather is ignored. 300\,s visits are used in these simulations.
\label{fig:CadenceTimeDiff}}
\end{figure}

As verification for the scheduler performance,
Figure~\ref{fig:CadenceTimeDiff} shows the time-difference (cadence) between
pairs of visits in the slow (upper panel),
and high (lower panel) cadences.
This plot is based on simulated one-year observations
without weather. We assumed eight mounts are allocated for the fast cadence and four for the slow cadence.

\begin{figure}[h]
\centerline{\includegraphics[width=7.5cm]{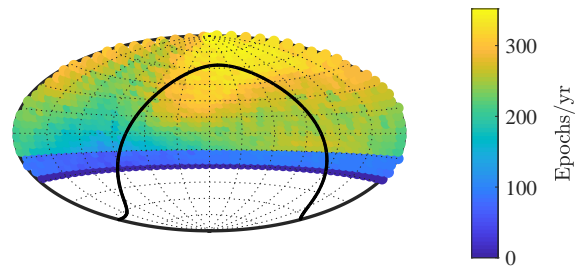}}
\caption{One year simulated number of visits per field, shown in Aitoff projection. The black line represents the Galactic equator. The difference in the number of visits between fields at the same declination zone is mainly due to variable length of night over the year.
\label{fig:VistisPerField}}
\end{figure}

\begin{figure}[h]
\centerline{\includegraphics[width=7.5cm]{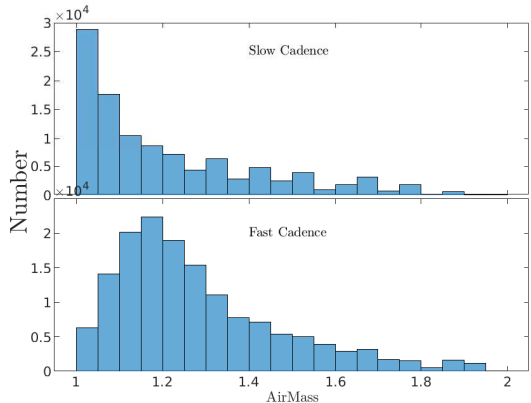}}
\caption{One year simulated visit's Hardie-airmass distribution in the slow (upper panel) and fast (lower panel) cadence surveys.
\label{fig:SimulatedAM}}
\end{figure}

For the same simulation, Figure~\ref{fig:VistisPerField} presents the number of visits per field for one year's worth of observations,
and Figure~\ref{fig:SimulatedAM} shows the expected airmass
distribution for the slow and fast cadence observations.
\vspace{3cm}

\subsubsection{Mount visibility and the allocator}
\label{subsubsec:alloc}

\begin{figure}
\centerline{\includegraphics[width=8cm]{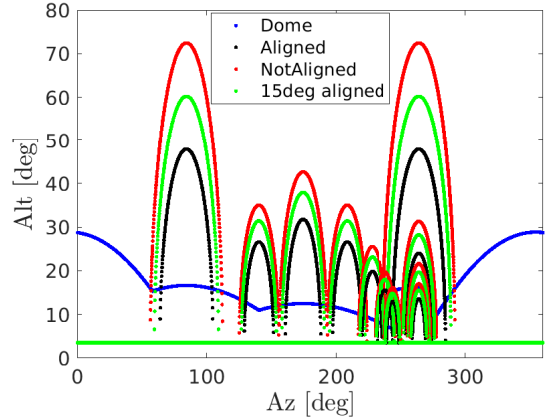}}
\caption{The altitude as a function of the azimuth of the horizon geometric obstruction model for mount number 3. Blue points show the obstruction due to the enclosure.
Black, green and red points are the obstruction due to the other telescopes in the array, for three possible configurations.
Black points for the case that all the telescopes are pointed in the same direction. Red points are for the worst-case random pointing of the telescopes, while green points are for the case in which the telescopes are co-aligned to the level of 15\,deg from each other.
\label{fig:Mount3_Obscuration}}
\end{figure}

\begin{figure}
\centerline{\includegraphics[width=8cm]{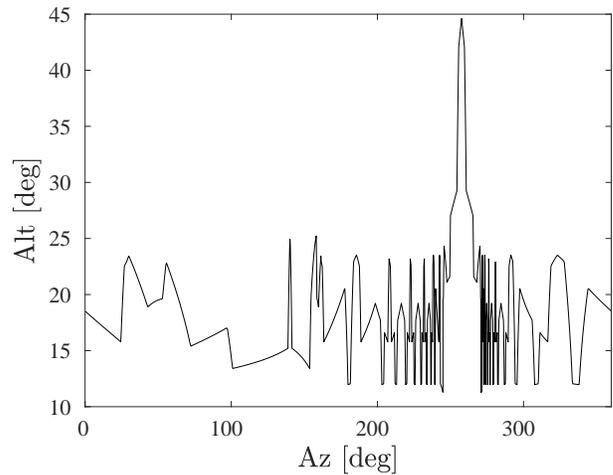}}
\caption{The minimum visible altitude as a function of azimuth, for at least one mount in the system.
This plot assumes that the telescopes are not aligned (i.e., the red curve in Figure~\ref{fig:Mount3_Obscuration}).
\label{fig:ObscurationMode_MinAll}}
\end{figure}

As the telescopes are packed closely inside the enclosure,
their horizon is limited by the enclosure walls, and the other telescopes in the enclosure.
The obstruction by other telescopes is complex, as it depends 
on the pointing direction of the telescopes.
In order to deal with this problem, for each telescope we have an
horizon obstruction model that depends on the pointing of other telescopes.
For example, in Figure~\ref{fig:Mount3_Obscuration} we show the obstruction model for the telescopes on mount number 3.
Blue points show the obstruction due to the enclosure.
Black, green and red points are the obstruction due to the other telescopes in the array, for three possible configurations.
Black points for the case that all the telescopes are pointed in the same direction. Red points are for the worst-case random pointing of the telescopes, while green points are for the case in which the telescopes are co-aligned to the level of 15\,deg from each other.
In Figure~\ref{fig:ObscurationMode_MinAll} we present the minimum visible altitude as a function of azimuth, for at least one mount in the system.
This plot assumes that the telescopes are not aligned (i.e., the red curve in Figure~\ref{fig:Mount3_Obscuration}).

Given these restrictions, each LAST mount can access between 76\% and 83\% of the sky above 30\,deg above the horizon.
However, 99\% of the sky above an altitude of 25\,deg is visible
for at least one mount.

The allocation of mounts to targets requires taking into account the sky visibility of each mount.
E.g., by first allocating the targets that can be observed by the smallest number of mounts.
In the rare cases in which the target cannot be allocated to a mount ($\lesssim 0.01$ of the cases),
a backup field is used.

\subsection{Apparatus calibration software}
\label{subsec:calib}

Given the large number of telescopes in the system, any alignment task (e.g., polar alignment) has to be done many times.
Therefore, we have developed automatic tools to perform the apparatus calibrations,
including:
polar alignment routines (\S\ref{subsec:Polar}),
focus and tip-tilt (\S\ref{subsec:FocusTipTilt}),
dark (\S\ref{subsec:dark}) and flat images (\S\ref{subsec:flat}),
and pointing model (\S\ref{subsec:PointingModel}).

\subsubsection{Polar alignment}
\label{subsec:Polar}

We have developed two automatic routines for polar alignment.
One is based on the popular drift method (e.g., \citealt{Tatum1978_PolarAlignMethods}),
and the second is based on observations of the polar region at different hour angles.
Both routines are fully automatic and report to the user, how much the mount hour angle axis should be moved.
However, since the second method is considerably faster (takes about 1 minute per iteration) it became our preferred method.

The second method uses the following steps:
(i) We set the telescope declination to 90 deg, and observe at several (at least 3)
hour angle positions with at least 120\,deg span;
(ii) For each image, we solve the astrometry (see pipeline paper and \citealt{Ofek2019_Astrometry_Code}),
and calculate the detector pixel position of the celestial pole, and some fiducial coordinate (e.g., Polaris);
(iii) A circle is fitted to the positions of the fiducial coordinates in the images taken at different hour angles, and the center of the best-fit circle is the estimated sky position towards the mount axis is pointing;
(iv) the offset of the mount axis from the celestial pole in azimuth and altitude is calculated and reported to the operator;
(v) The operator is responsible for shifting the mount hour-angle axis in azimuth and altitude.
Typically, four to five iterations of this process are enough in order to complete the polar alignment
to an accuracy of about one arcminute.

\subsubsection{Focus and tip-tilt}
\label{subsec:FocusTipTilt}

The LAST telescopes are focused by moving the primary mirror.
However, additional degrees of freedom exist.
Specifically, the tip-tilt of the field flattener, relative to the corrector plate, can be controlled via three screws.
In order to simplify this process,
we added the capability to control the piston tip and tilt of the camera,
with respect to the field flattener by the insertion of shims into the camera-to-telescope adapter.

Focusing is performed by looping through focus values around some nominal focus position.
In each focus position, an image is taken and analyzed.
We used two kinds of analysis, one is adequate when the telescope
is near focus and the other when the telescope is far away from focus.
We first execute a source-finding routine using a large iterative template bank.
The source-finding routine includes cross-correlating
(filtering) the background-subtracted image,
normalizing it by the filtered image standard deviation (StD; e.g., \citealt{Zackay+2017_CoadditionI}), and searching
for local maxima above $S/N$ of 50.
This process is performed (simultaneously) using multiple
Gaussian filters with different widths.
For each source, we mark the filter that maximizes 
the source $S/N$.
The filter which has the largest number of maximal sources $S/N$ is declared as the
the best fit seeing.
To expedite this process, it is done iteratively.
In the first iteration, we start with five templates logarithmically spaced between 0.6 to 100 pixels sigma-width. 
Next, the best template is chosen,
and we repeat this process, this time around the best
template with a range defined by the templates
next to the best template (one below and one above).
Typically, it takes about six iterations for this
process to converge.
To save time, this routine is performed for a region of about 2000 by 2000 pixels around the detector center.
However, if the telescope is far away from focus this routine will fail,
and instead, we subtract the background and set to zero pixels with a value that is smaller than $5\sigma$ above the background noise.
Next, we calculate the auto-correlation function of this image and measure the width of the central peak of the
auto-correlation image.
This gives us an estimate of the FWHM of the sources in the image.

To find the tip/tilt of the camera,
we execute the focus routine, but for a grid of positions on the detector plane 
(example in \S\ref{subsec:ImageQuality}), and fit a best focus parabolic surface to the focus value as a function
of position. 
The parabolic surface has the form of
\begin{equation}
    F(x,y) = a_{1} + a_{2}x + a_{3}y + a_{4}xy + a_{5}x^{2} + a_{6}y^{2}.
\end{equation}
Here $F(x,y)$ is the PSF FWHM as a function of $x$ and $y$ pixel position.
The $a_{2}$ and $a_{3}$ terms represent the tip and tilt of the camera, and our optimization calls for finding the tip-tilt that minimizes these terms.
Next, we find the piston that minimizes
the FWHM.

\subsubsection{Dark images}
\label{subsec:dark}

Since the telescopes are not equipped with
remotely-operated covers,
the dark images are taken during system maintenance.
20 dark images are taken at
the nominal operating temperature of the
camera, of about $-5^{\circ}$C.
A master dark, variance map, and 32-bit pixel mask image are generated (see pipeline paper for details).
The mask image contains information
about pixels with low StD, high StD, flaring pixels, and pixels with high values.

\subsubsection{Flat fielding}
\label{subsec:flat}

A flat-fielding system is under consideration,
but currently, we are using twilight flat-fielding.
The flat script operates
during twilight when the Sun's altitude
is between $-3$\,deg to $-8$\,deg.
During this time, the sky brightness
is measured in one-second exposures,
and the exposure time needed to produce
a mean count between 6000 and 40000 is estimated.
If this exposure time is between 3\,s
to 20\,s then a flat image is taken.
During such a session, about 15 flat images
are obtained.
The telescopes are moved randomly by about 1\,deg,
between flat images.
Furthermore, the pointing of the telescope during twilight flat is selected
to be near the zenith and avoids, if possible, the Galactic plane and the Moon.
%\textbf{GN: I would be more worried about the ecliptic/planets in the flat field.}

The flat images are used in order to generate a master
flat image along with a variance image,
and a 32-bit per pixel mask image.
The mask image flags pixels with large variance,
low response, or a NaN value
(see \citealt{Ofek+2023ASP_LAST_PipelineI} for details).

\subsubsection{Pointing Model}
\label{subsec:PointingModel}

In order to improve the mounts' pointing
we map the difference between each
mount's encoder-based coordinates and the true
astrometric pointing of the four telescopes on the mount.
%This procedure takes into account the
%Earth nutation
%and the UT1$-$UTC time difference.

This procedure is conducted in two modes.
The first mode is executed in the first
observing run after the polar alignment step.
This includes observing about 100 pointings over the entire celestial sphere with airmass smaller than about 2.
For each pointing we solve the astrometry
and store the difference between the encoder
coordinates and astrometric coordinates
in the mount coordinates correction table stored in a configuration file.
The pointing errors are interpolated from this table.

A second procedure uses all the scientific data
to monitor and refine, if needed, the coordinates pointing model table.
Our 20-exposures per visit strategy also offers information about tracking errors.
For each set of 20 exposures the linear tracking error is logged in the coadd image headers, and if needed a model describing
the deviations in the tracking
from the sidereal rate can be constructed,
and applied to the telescope tracking rate.

\section{Camera performance}
\label{sec:Cam}

Here we describe the performances of the
SONY IMX455/QHY600M-PH camera.
An independent analysis of the characterization
of this camera was recently published in \cite{Alarcon+2023_CMOS_QHY600_QHY411_characterization}.

\subsection{Quantum Efficiency}
\label{subsec:QE}

In order to measure the IMX455 quantum efficiency,
we have used a double monochromator setup to produce wavelength-tunable monochromatic light between UV and IR wavelengths. To account for temporal changes in the light source flux (\citealt{Kusters+2020SPIE_LightSorce_Calibration,Kusters+2022SPIE_LightSourceCalibration}) we illuminated the detector (including the QHY600 window) under test in parallel to a reference diode using a beam splitter. For precisely synchronized detectors we can reach $<0.1$\% accuracy. As we do not exactly know the precise timing of the QHY600 exposure we are limited to the variations in the laser-driven light source which are of the order of 1\% in flux. Improvements would be possible with the use of a common shutter for all detectors.
The measurement of the quantum efficiency is then a two-step process. We first place a NIST traceable detector at the measurement beam and calibrate the reference diode, afterwards we place the QHY600 in that beam and repeat the measurement. 

\begin{figure}
\centerline{\includegraphics[width=8cm]{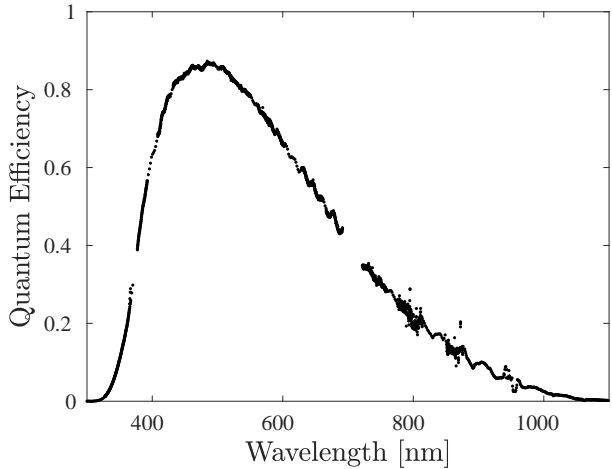}}
\caption{The quantum efficiency of the IMX455 detector, including the QHY600M-PH window
(see Table~\ref{tab:QHY600_QE}).
\label{fig:QHY600_QE}}
\end{figure}

To calibrate the wavelength scale we measure a Holmium Didymium (HoDi) absorption line filter with each measurement, in the calibration of the reference diode as well as in the measurements with the QHY600. In this way we imprint the absorption lines of the HoDi filter to the output spectrum of our light source. The position of the HoDi absorption lines was calibrated in advance, against the emission lines of low-pressure gas lamps. The emission line positions are then taken from the NIST atomic spectra database lines form\footnote{https://physics.nist.gov/PhysRefData/ASD/lines\_form.html}.
The measured quantum efficiency is shown in Figure~\ref{fig:QHY600_QE},
and it is listed in Table~\ref{tab:QHY600_QE}.
\begin{deluxetable}{lrl}
\tablecolumns{3}
\tablewidth{0pt}
%\tabletypesize{\footnotesize}
\tablecaption{Quantum Efficiency of the IMX455 detector and window}
\tablehead{
\colhead{Wavelength}    &
\colhead{Efficiency}    &
\colhead{Uncertainty}   \\
\colhead{[nm]}       &
\colhead{}           &
\colhead{} 
}
\startdata
299.97 & $-0.00006$ & 0.00041 \\
301.55 & $ 0.00032$ & 0.00037 \\
303.15 & $ 0.00026$ & 0.00034 \\
304.72 & $ 0.00017$ & 0.00031 \\
306.29 & $-0.00011$ & 0.00028
\enddata
\tablecomments{The full table is given in the electronic version of the paper. Here we list the first five entries.}
\label{tab:QHY600_QE}
\end{deluxetable}

\subsection{Camera parameters}
\label{subsec:Gain}

\begin{figure}
\centerline{\includegraphics[width=8cm]{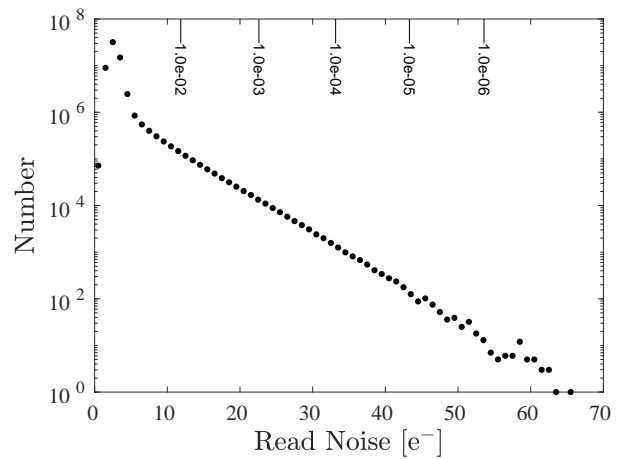}}
\caption{The distribution of the QHY camera readout noise per pixel (as measured for camera QHY600M-51a1fd4485a353a51), for mode 1, gain parameter 0 and offset 6. Single electron bins are used. The upper axis shows one minus the cumulative fraction of pixels with the given read-noise values.
\label{fig:ReadNoise_PerPix_QHY600}}
\end{figure}

The camera can be operated in several modes,
and in each mode, the gain and bias level can be controlled.
Our selected default setup, for which parameters are listed in Table~\ref{tab:cameraPars}, is a compromise between readout noise and dynamic range.
The camera gain parameter (specified in arbitrary units) controls the actual gain value and readout noise.
%We are using the gain parameter that maximizes the dynamic range of the camera.
The camera offset parameter (bias) was selected such that it is the lowest possible while the number of pixels with counts of $<3$\,ADU is below 50.
In Table~\ref{tab:cameraPars} we list the
parameters we have measured for the selected gain.
%These parameters are similar to those given in the manufacturer website\footnote{}.
To measure the read-out noise we obtained ten bias frames and measured the StD over each pixel,
and took the mean or median of all the StD values.
Since for many types of measurements we are interested
in the total counts in several adjacent pixels
(e.g., aperture photometry), we also convolved the image of
StD per pixel with a $3\times3$ and $5\times5$ top-hat square, and list the mean and median readout noise in such $3\times3$ and $5\times5$ apertures.
We note that the measured read-out noise
we report in Table~\ref{tab:cameraPars}
is a lower than the readout noise
reported in \cite{Alarcon+2023_CMOS_QHY600_QHY411_characterization} ($3.48$\,e$^{-}$)
or the manufacturer ($3.67$\,e$^{-}$).
\begin{deluxetable}{ll}
\tablecolumns{2}
\tablewidth{0pt}
%\tabletypesize{\footnotesize}
\tablecaption{QHY600M-PH camera parameters}
\tablehead{
\colhead{Property}    &
\colhead{Value}      \\
\colhead{}       &
\colhead{}  
}
\startdata
\hline
Mode              & 1   \\
Gain parameter    & 0   \\
Offset            & 6   \\
\hline
Gain                        & 0.75\,e$^{-}$/ADU\\
Mean Readout noise $1\times1$ & 3.0\,e$^{-}$ \\
Median Readout noise $1\times1$ & 2.7\,e$^{-}$\\
Mean Readout noise $3\times3$ & 3.0\,e$^{-}$\\
Median Readout noise $3\times3$ & 2.8\,e$^{-}$  \\
Mean Readout noise $5\times5$ & 3.0\,e$^{-}$\\
Median Readout noise $5\times5$ & 2.9\,e$^{-}$\\
Dark current at $-5$C & $8\times10^{-3}$\,e$^{-}$\,pix$^{-1}$\,s$^{-1}$ \\
Bias level       & $\approx99$\,ADU  \\
Full well        & $\approx48,000$\,e$^{-}$
\enddata
\tablecomments{The QHY600M-PH camera setup mode and measured parameters.}
\label{tab:cameraPars}
\end{deluxetable}
Figure~\ref{fig:ReadNoise_PerPix_QHY600} presents the histogram of the readout noise
per pixel
%

%Also measured are the bias level and dark current at our nominal working temperature, and they are listed in Table~\ref{tab:camera}.

\subsection{Non-linearity}
\label{subsec:NonLin}

\begin{figure}
\centerline{\includegraphics[width=8cm]{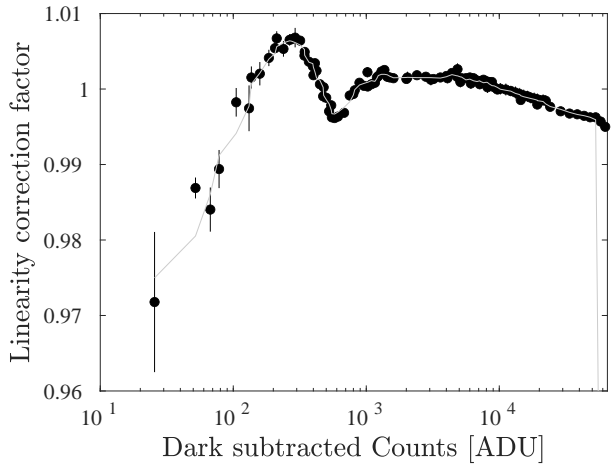}}
\caption{The non-linearity correction factor (for camera QHY600M-51a1fd4485a353a51), as a function of counts in ADU. The gray line shows a Savitzky-Golay (first order, length of 5) smoothed version of the data.
\label{fig:LinearityQHY600}}
\end{figure}

The camera's non-linearity, at the selected mode and gain parameters, was measured as follows:
We use an intensity-stabilized light source fed into a collimator via a fiber link.
The collimated beam was projected onto the detector via a set of neutral-density filters.
We obtained images as a function of exposure time, subtracted the dark image, and measured the mean count rate
in a sub-image of size 100 by 100 pixels.
The median of each sub-image was divided by its exposure time.
This resulted in the linearity correction factor as a function of the counts in ADU.
To cover the entire dynamic range of the camera, we repeated this with different neutral-density filters.
The sets of measurements, obtained using different filters, were calibrated
to the same correction factor value using their overlapping regions.
Next, we normalized the correction factor as a function of ADU
such that the correction factor at 10,000\,ADU will be exactly 1.
In order to estimate the errors
and stability,
we repeated these measurements five times,
and in each count level, we calculated the standard deviation of the measurements
divided by $\sqrt{5}$.
Finally, we compare this stability
estimate with the error calculated from
the Poisson errors.
The two measurements of the uncertainty agree well.
In Figure~\ref{fig:LinearityQHY600} we show the non-linearity correction factor (response)
for one of our cameras as a function of the counts in ADU.
The largest deviations from non-linearity appear at a low count rate,
while for intermediate and high counts the non-linearity correction is below 1\%.
Testing different cameras, we find that 
there are small but statistically significant, differences
between different cameras.
However, the differences between cameras are small:
Typically, less than 1\% at very low counts of about $<50$\,ADU above bias (typically well within the sky level), and about
$\sim0.1\%$ at a higher count rate.
In the pipeline (\citealt{Ofek+2023ASP_LAST_PipelineI}) we correct for the non-linearity
using a smoothed version of these measurements (gray line in Fig.~\ref{fig:LinearityQHY600}).

\section{On sky performances}
\label{sec:Perform}

Here we present some of the on-sky performance of LAST,
including vignetting (\S\ref{subsec:vig}),
image quality (\S\ref{subsec:ImageQuality}), and
limiting magnitude (\S\ref{subsec:LimMag}).
%and some initial science results (\S\ref{subsec:ScienceRes}).
%
Additional aspects, related to the LAST pipeline,
as photometric and astrometric precision, are presented
in \cite{Ofek+2023ASP_LAST_PipelineI}, and are summarized in Table~\ref{tab:perf}.
\begin{deluxetable}{ll}
\tablecolumns{2}
\tablewidth{0pt}
%\tabletypesize{\footnotesize}
\tablecaption{Photometry and Astrometric performances}
\tablehead{
\colhead{Property}    &
\colhead{Value}      \\
\colhead{}       &
\colhead{}  
}
\startdata
Median zero point accuracy  & 0.01\,mag \\
Relative photometry precision in 20\,s (over 4000\,s)  & 0.004\,mag \\
Relative photometry precision in 320\,s (over 4000\,s)  & 0.001\,mag \\
\hline
Astrometric accuracy (20\,s) & 60\,mas \\
Astrometric accuracy (coadd $20\times20$\,s) & 30\,mas\\
Astrometric accuracy (averaged over $16\times20$\,s) & 15\,mas
\enddata
\tablecomments{A summary of LAST photometric and astrometric precision.
All the numbers are for the bright end.
For more details see the pipeline paper.
Zero point accuracy is measured relative to the reference catalog GAIA-DR3 (\citealt{GAIA+2021_GAIAEDR3_Summary_Content}).}
\label{tab:perf}
\end{deluxetable}

\subsection{Vignetting}
\label{subsec:vig}

\begin{figure}
\centerline{\includegraphics[width=8cm]{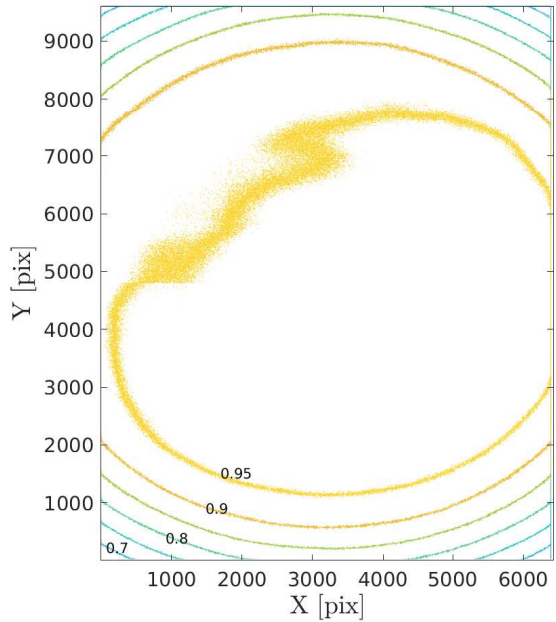}}
\caption{The typical vignetting pattern of a LAST telescope, as measured from a flat field image (after a 4 by 4 median filter).
We assume that the pixel scale is uniform across the field of view.
The vignetting was normalized to have a maximum of 1.
Contour levels are shown in steps of $0.05$.
\label{fig:LAST_Vignetting}}
\end{figure}

LAST uses a modified Schmidt telescope
in which the mirror size is equal to the pupil size,
and it has a large obscuration due to the prime focus camera.
In Figure~\ref{fig:LAST_Vignetting} we present the typical vignetting pattern of a LAST telescope, as measured from a flat field image.
Normalizing the vignetting to have a peak value of 1,
about 72\% (93\%) of the LAST image area has a vignetting
$>0.9$ ($>0.8$).

\subsection{Image quality}
\label{subsec:ImageQuality}

\begin{figure}
\centerline{\includegraphics[width=8cm]{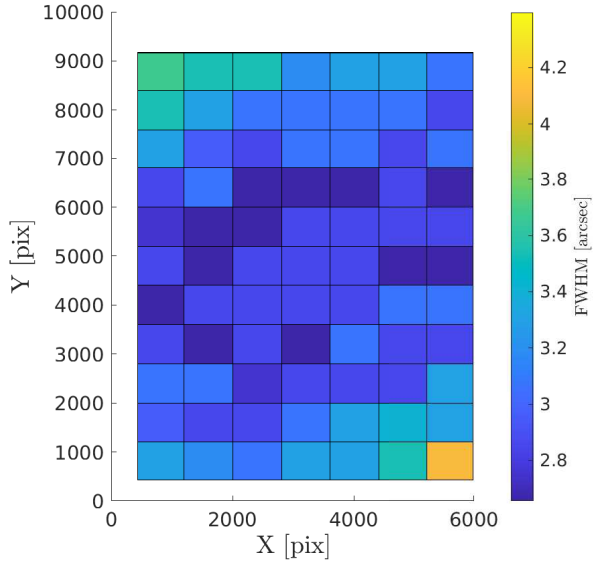}}
\caption{An example for the measured FWHM as a function of position on the detector.
The FWHM was measured in bins of 800 by 800 pixels.
In this example, the mean FWHM over the image is about $3.1''$.
\label{fig:ImageQuality}}
\end{figure}

\begin{figure*}
\begin{align}
\includegraphics[width=8.5cm]{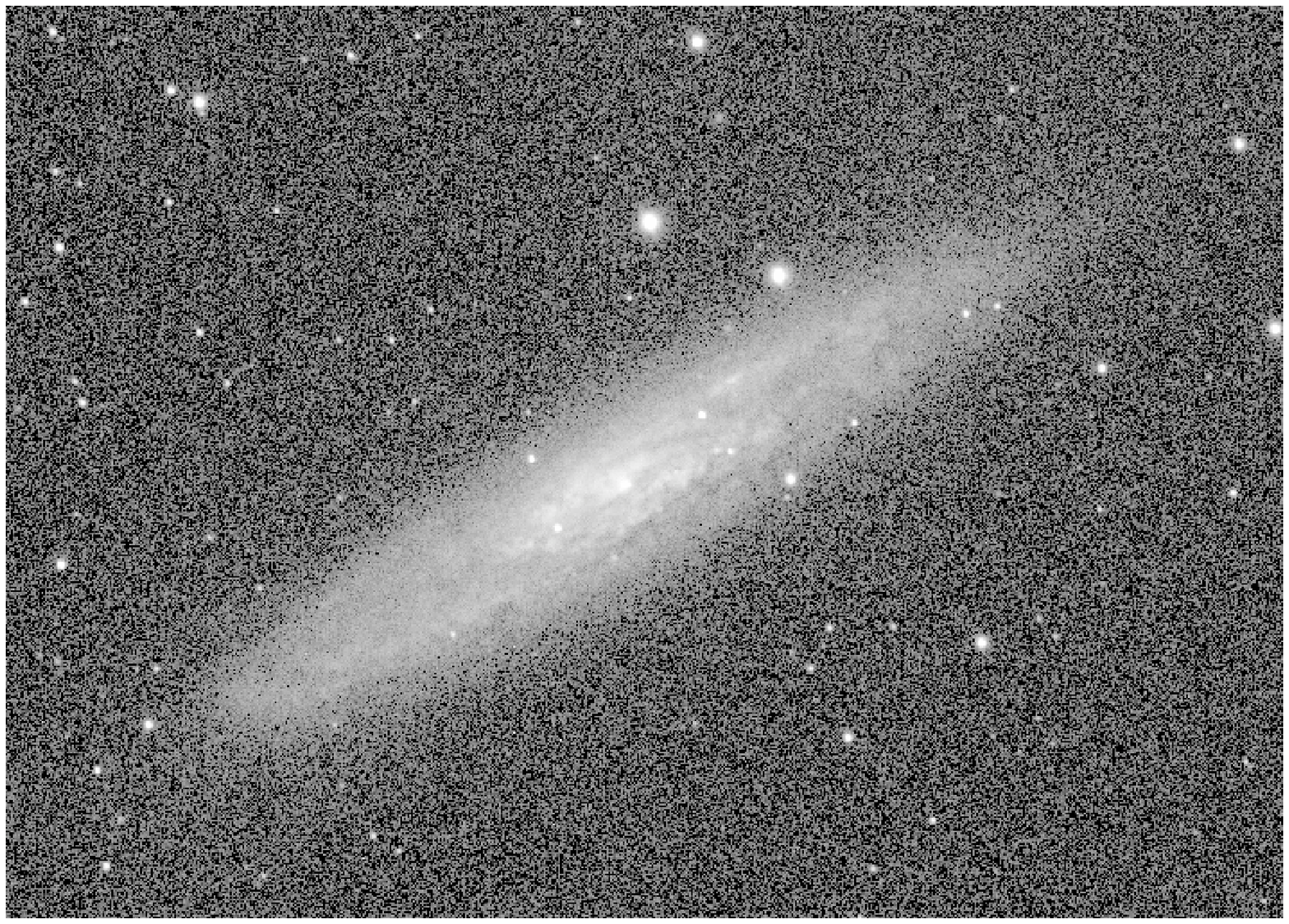} &
\includegraphics[width=8.5cm]{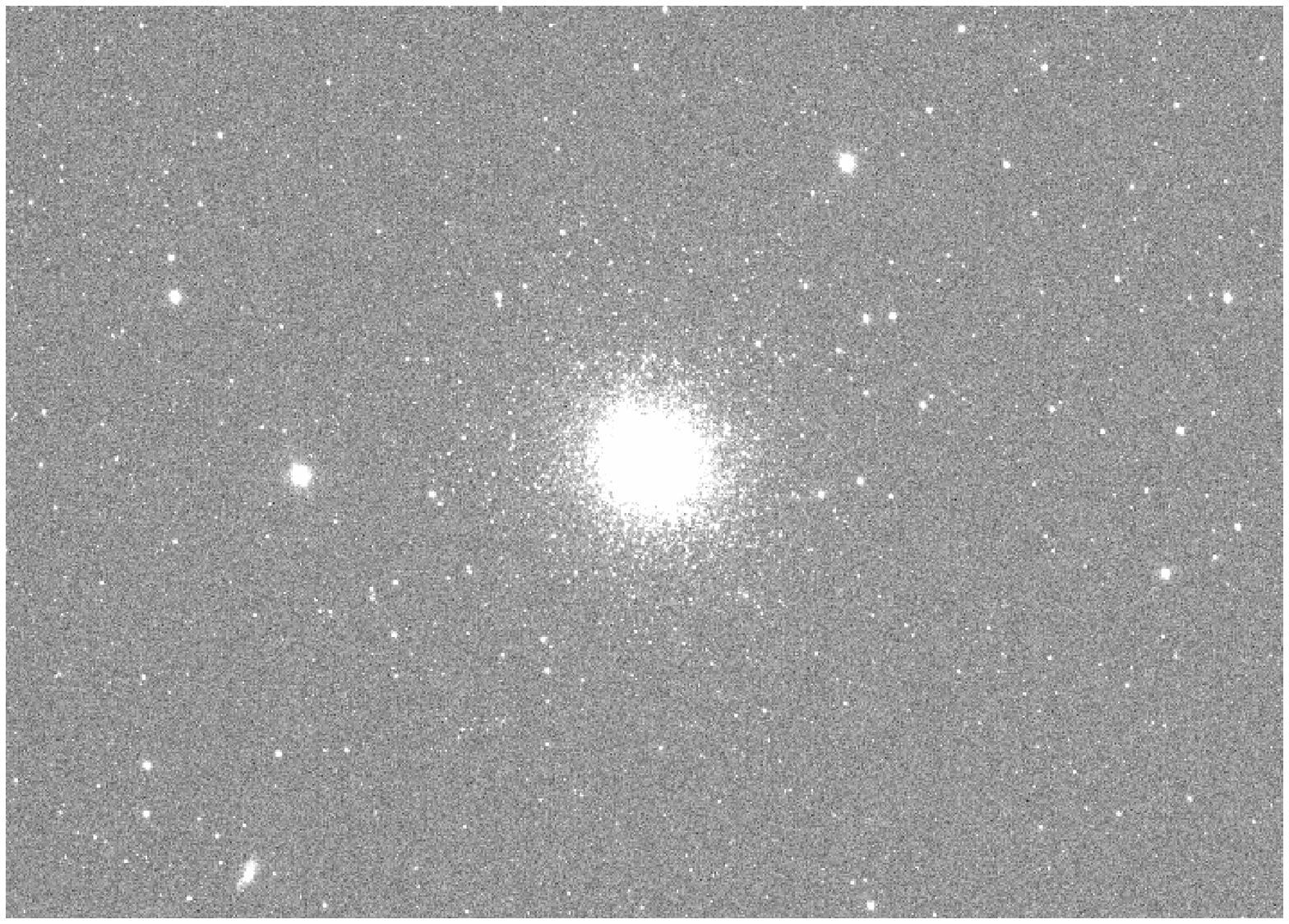} \\
\includegraphics[width=8.5cm]{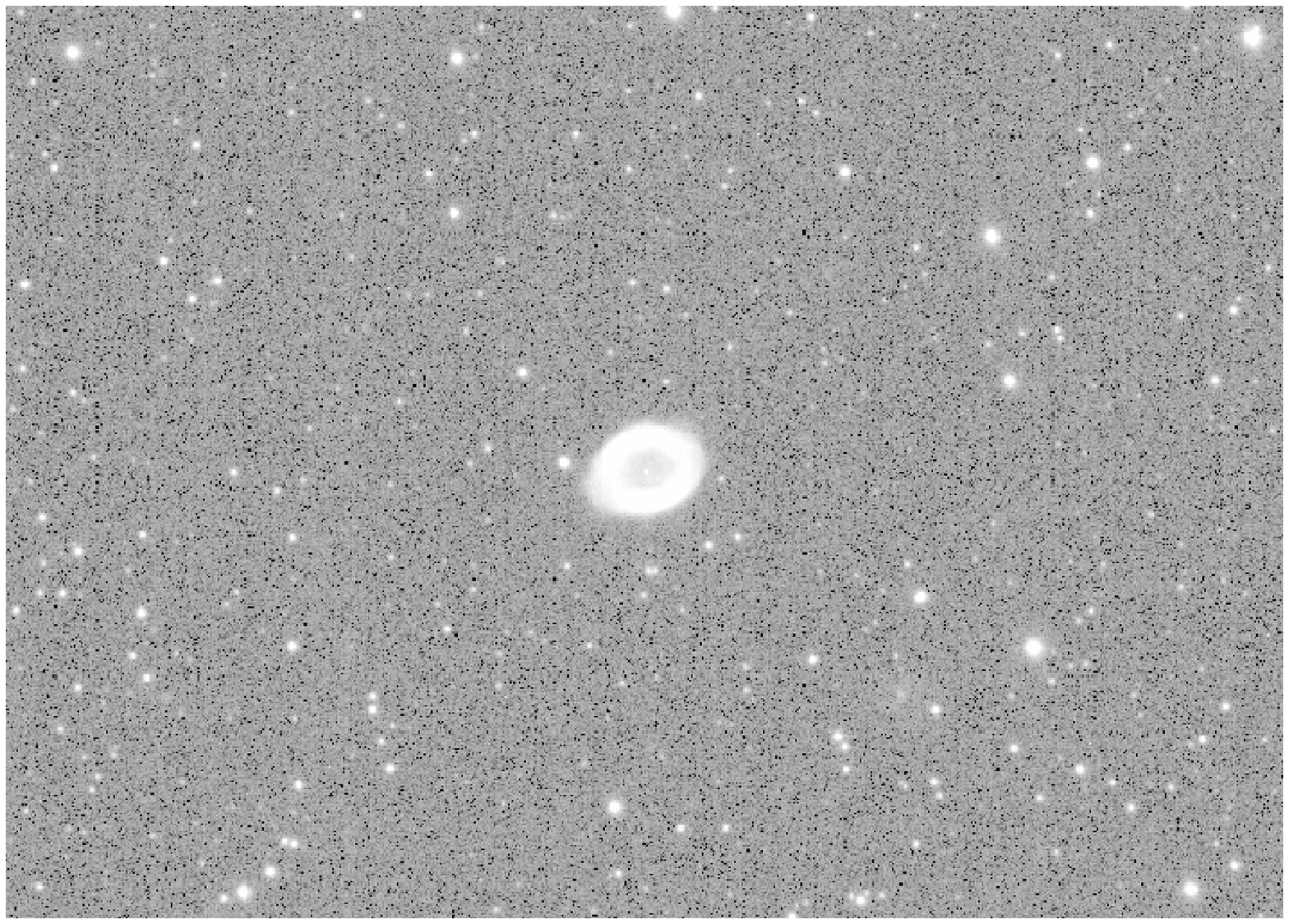} &
\includegraphics[width=8.5cm]{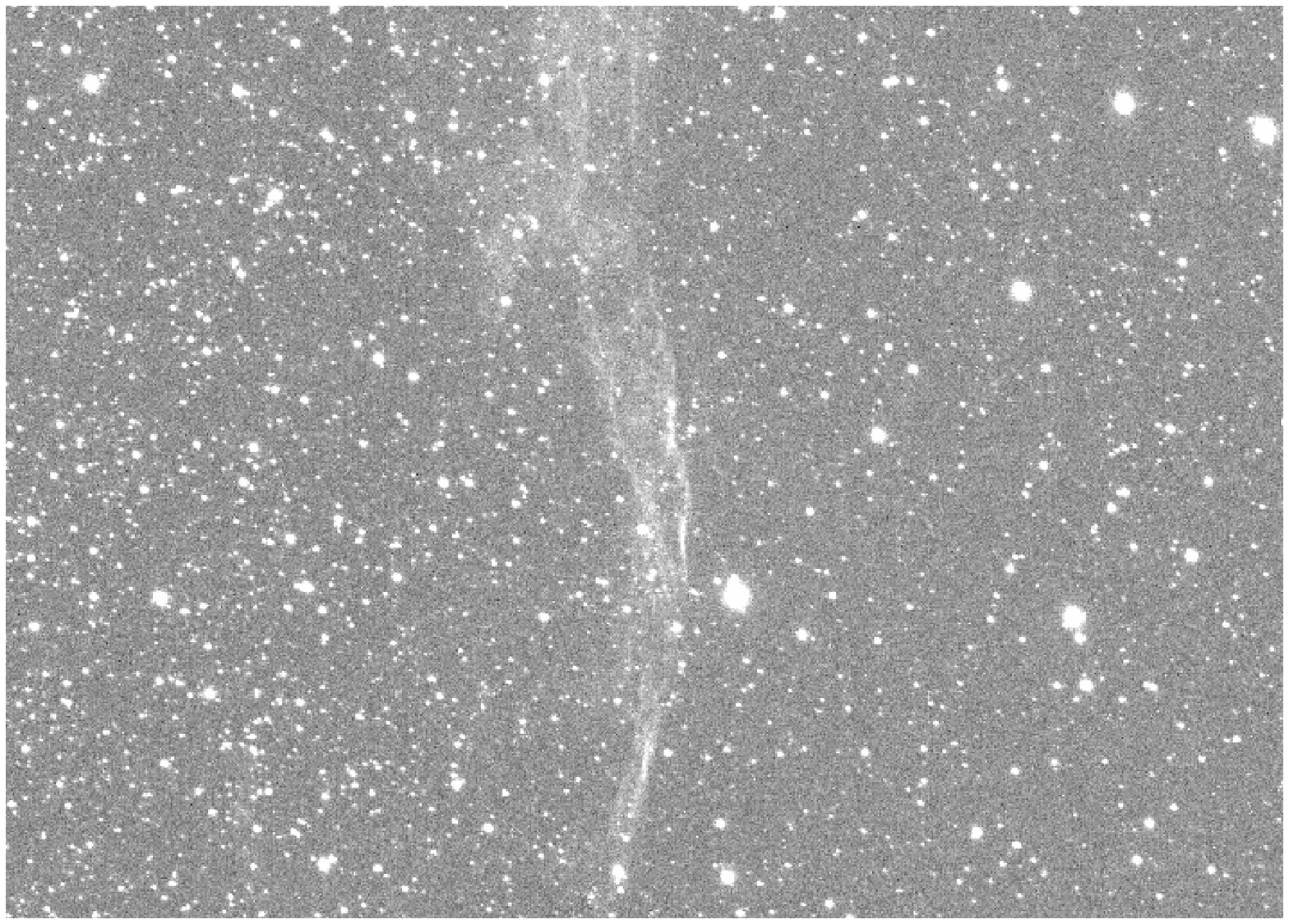} \\
\includegraphics[width=8.5cm]{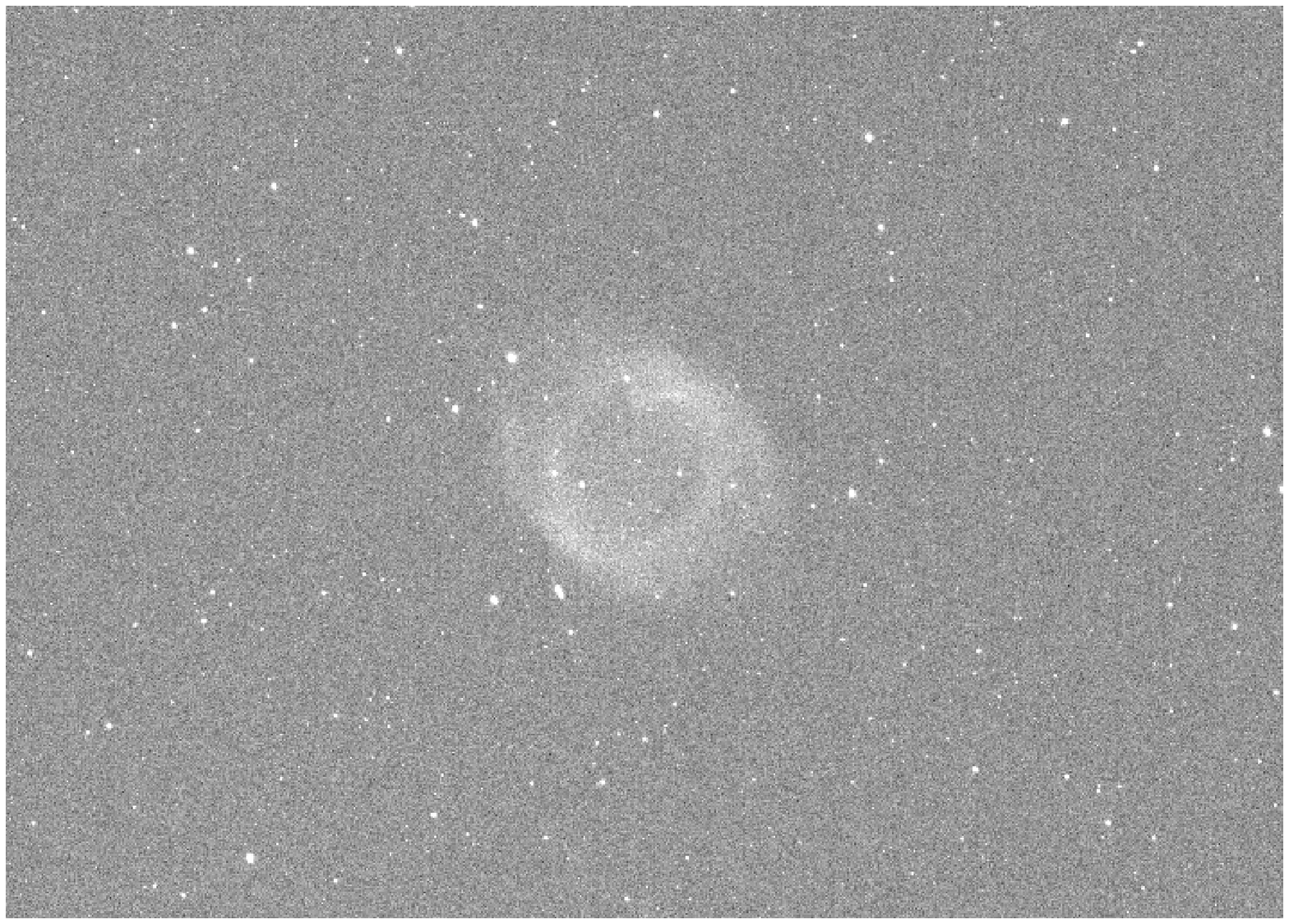} &
\includegraphics[width=8.5cm]{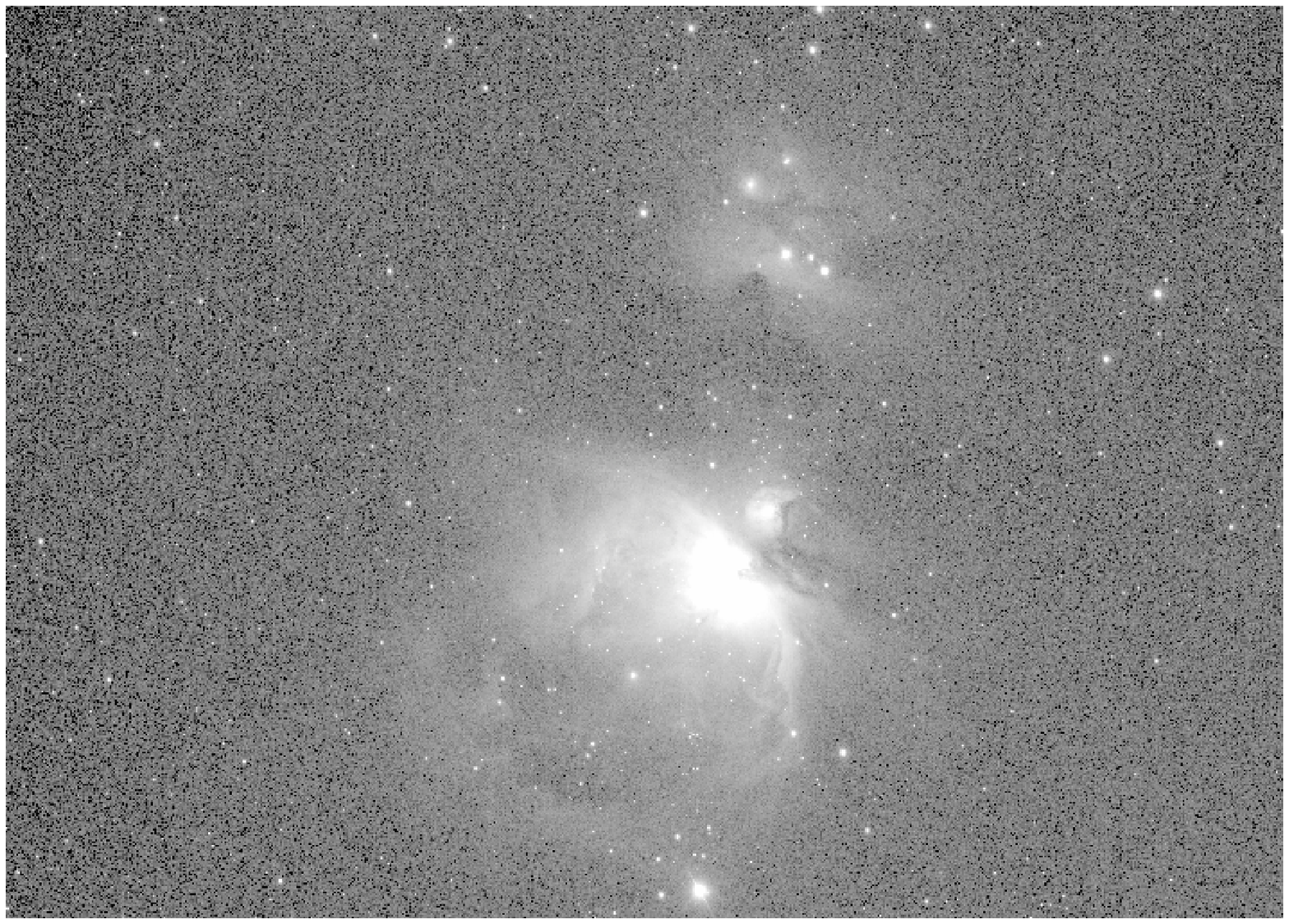} 
\end{align}
\caption{Small sections (out of the entire camera field of view) around selected objects.
Exposure times are 15\,s using a single telescope.
Left to right, top to bottom are:
NGC\,253, M13, M57, the Veil, the Helix, and M42.
NGC253 and M42 are presented in logarithmic scale.
\label{fig:LAST_NGC253}}
\end{figure*}

Before applying tip-tilt corrections to the cameras,
the measured FWHM, near the image center, ranges from $1.9''$ to $2.8''$.
An example of the measured image quality in one of the telescopes/detectors before we applied tip-tilt correction is shown in Figure~\ref{fig:ImageQuality}.
%
%Figure~\ref{} shows the FWHM as a function of {\bf add figure} position
%on the detector, for a single image with good image quality.
Given the site median seeing of $1.4''$,
the pixel scale of $1.25''$\,pixel$^{-1}$,
and the theoretical delivered image quality
these are reasonable performance results.
Figure~\,\ref{fig:LAST_NGC253} shows several
15\,s image cutouts around selected objects. 

\subsection{Limiting Magnitude}
\label{subsec:LimMag}

\begin{figure}
\centerline{\includegraphics[width=8cm]{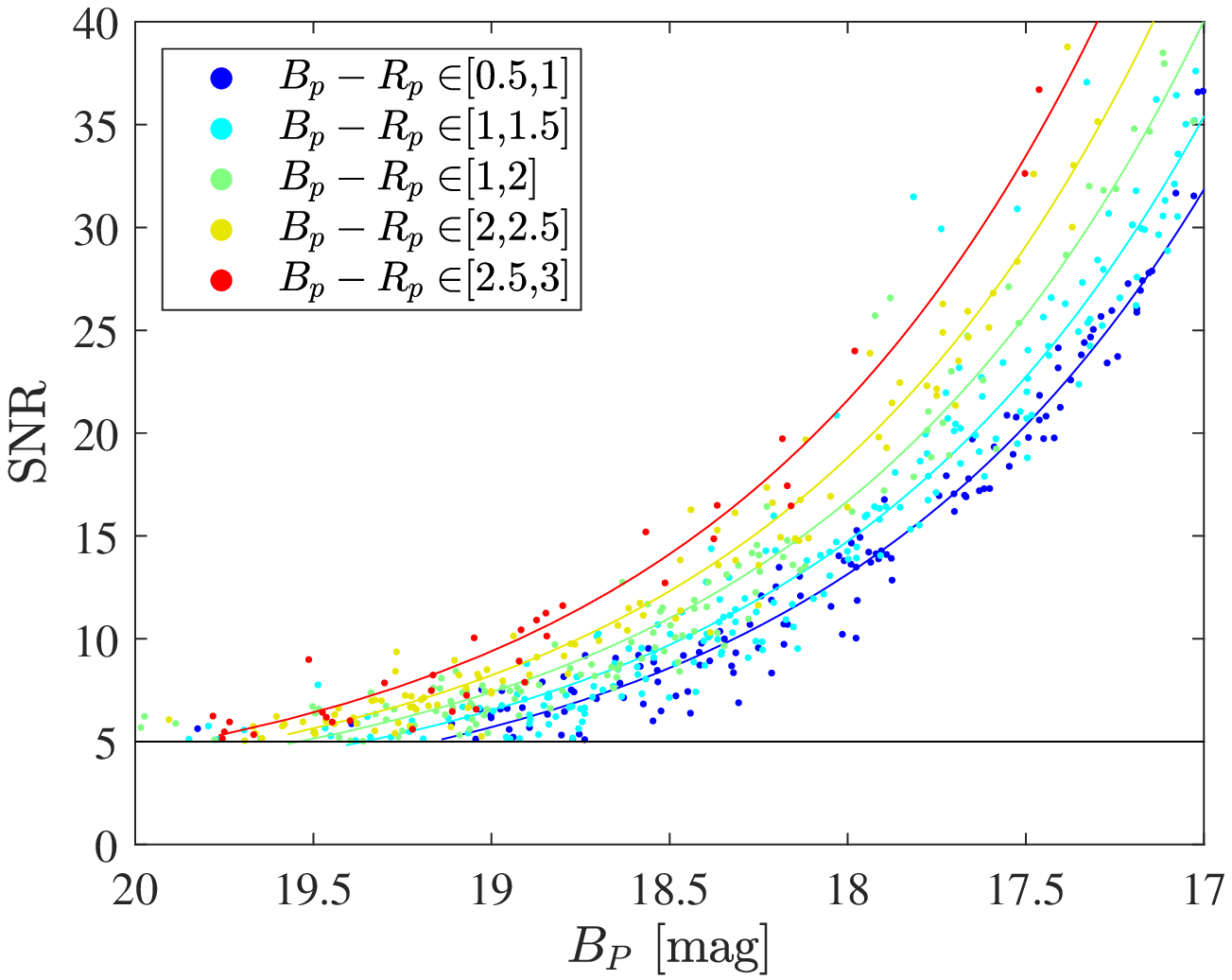}}
\caption{The Signal-to-Noise ratio ($S/N$) as a function of the $B_{\rm p}$ AB magnitude, for stars with different colors in a 15\,s exposure (dots; color coding: see legend). The colored lines show the best fit $\log_{10}(S/N)$ vs. $B_{\rm p}$ AB magnitude to stars in different color ranges. The limiting magnitude is estimated by reading the value of the fitted line at $S/N=5$ and color equal 1.
\label{fig:SN_Bp}}
\end{figure}

Out of the three GAIA bands (\citealt{GAIA+2016_GAIA_mission}),
the IMX455 sensitivity has the highest resemblance to the GAIA $B_{\rm p}$ band. Therefore, our calibration is done relative to this band.
Here, we convert all the GAIA magnitudes to the AB magnitude system. This is done by adding $0.1136$, $0.0155$, and $0.3561$\,magnitude, to the
$G$, $B_{\rm p}$, and $R_{\rm p}$ GAIA Vega magnitudes, respectively.
Figure~\ref{fig:SN_Bp} shows the ${S/N}$ for detection vs. the GAIA $B_{\rm p}$ magnitude for one representative (15\,s) image taken at dark time near the zenith.
Points with different colors represent
sources with different $B_{\rm p}-R_{\rm p}$ color
(see legend).
The colored lines show the logarithmic best-fit lines 
for sources in different color bins.
The limiting magnitude for each image and coadd image
is calculated by linear fitting the $\log_{10}(S/N)$
as a function of the GAIA $B_{\rm p}$ magnitude
and the $B_{\rm p}-R_{\rm p}$ color.
Next, the 5-$\sigma$ limiting magnitude is calculated by
reading the value of the fitted function
at $S/N=5$ and $B_{\rm p}-R_{\rm p}=1.0$\,mag.

The typical, dark time, 5-$\sigma$ limiting magnitude
in a 20\,s exposure is about 19.6 (for $V$-band sky brightness of 21\,mag\,arcsec$^{-2}$).
Using the coaddition of $20\times20$\,s exposures
the 5-$\sigma$ limiting magnitude,
for $B_{\rm p}-R_{\rm p}=1.0$\,mag, is about 21.0.
The system's limiting magnitude is color dependent,
with a slope of $+0.33$\,mag per $B_{\rm p}-R_{\rm p}$ magnitudes.
We estimate that by building a LAST node in a dark-site location
may improve its limiting magnitude by about 0.5\,mag.

\section{Future followup facilities}
\label{sec:FollowUp}

Only about 10\% 
of the transients currently found by sky surveys are being followed up spectroscopically
(\citealt{Kulkarni2020arXiv_FollowupOpticalTransients}).
This is a big limiting factor that is likely to become even more problematic in the near future.
Furthermore, for some applications, multi-band photometric observation is valuable for studying the objects of interest (e.g., for measuring bolometric light curves).
However, LAST is designed as a discovery machine, and
therefore it is normally not equipped with filters.

For these reasons we designed two new follow-up facilities, one for photometric observations,
and the second for spectroscopic follow-up.
A major design goal of these new facilities is, again, cost-effectiveness.

The Pan-chromatic Array for Survey Telescopes (PAST) is a planned photometric telescope.
The PAST design calls for four 14-inch f/11 telescopes on a single mount. Each telescope will be equipped with a dichroic filter and two cameras. The cameras will be equipped with broad-band filters (1500 to 2000\,\AA~width). The eight filters on each mount will cover the 4000\,\AA~to 8000\,\AA~range with large overlaps, such that their linear combination will allow us to get photometry in 500\,\AA~bands (see Ofek \& Zackay, in prep.).
This will allow us, on one hand, to get deep, broad-band imaging, and on the other hand to obtain some spectral information.
With the exception of the telescopes and dichroic filters,
the PAST components are identical to those used in LAST. 

The Multi-Aperture Spectroscopic Telescope (MAST; Ben-Ami et al., in prep.) will use a large number of small telescopes to collect the light from a target into a single spectrograph.
Our goal is to be able to construct a telescope with a collecting area equivalent to a 2.7\,m telescope, but for about 10\% of the price of such a telescope.

\section{Conclusions}
\label{sec:Conclusions}

We present the LAST project -- A cost-effective high-grasp
survey telescope for exploring the variable and transient sky.
The first LAST node of 48, 28\,cm telescopes,
providing a total field of view of 355\,deg$^{2}$,
is currently under construction.
The first 12 telescopes saw their first light in early March 2022, while additional 20 telescopes
will be installed on March 2023, and the rest are expected to be deployed by June 2023.
The development and construction time of LAST was relatively short -- About three years from our first experimentation
with the telescopes and some previous-generation
camera and mounts we tested, to first light.

The LAST strategy of obtaining several images (the default is 20)
of each field visit allows us, among other things, to identify
minor planets, satellite glints, white dwarf transits, and flare stars (\citealt{Ben-Ami+2023PASP_LAST_Science}).
In turn, this provides a cleaner stream for transient detection.
The LAST survey strategy will concentrate on a high cadence survey of the sky -- this
has the potential to open a new window into the fast transients and variability phase space
(e.g., \citealt{Drout+2014_Rapidly_Evolving, Ho+2021_ZTF_RapidlyEvolvingTransients, Ofek+2021_AT2018lqh_RapidlyEvolving}).
Some initial science results from LAST are presented in Ofek et al. (submitted).

An important aspect of the LAST design is its cost-effectiveness.
In terms of volume of space per unit time per unit cost (i.e., grasp per unit cost),
LAST is an order of magnitude improvement over most existing surveys.
We believe that this point is important, because, it provides a path to construct
affordable, very high grasp systems, around the globe.
Such systems are needed in order to provide continuous monitoring of the sky and to probe the fast transients and variability phase space.
Furthermore, with a large number of these systems,
their collecting area will become competitive
with existing large telescopes.
We emphasize, that this approach of multiple small telescopes can compete with large telescopes for seeing-limited observations, and likely also seeing-limited single-object spectroscopy (\S\ref{sec:FollowUp}; \citealt{Ofek+BenAmi2020_Grasp_SkySurvrys_CostEffectivness}).
However, large telescopes still have major advantages when equipped with diffraction-limited adaptive optics, $K$-band observations, multi-object spectroscopy, and additional unique instrumentation.

\acknowledgments

E.O.O. is grateful for the support of
grants from the 
Willner Family Leadership Institute,
André Deloro Institute,
Paul and Tina Gardner,
The Norman E Alexander Family M Foundation ULTRASAT Data Center Fund,
Israel Science Foundation,
Israeli Ministry of Science,
Minerva,
BSF, BSF-transformative, NSF-BSF,
Israel Council for Higher Education (VATAT),
Sagol Weizmann-MIT,
Yeda-Sela,
Weizmann-UK,
the Rosa and Emilio Segre Research Award,
Benozyio center,
and the Helen-Kimmel center.
A.F. and V.F.R. acknowledge support from the German Science Foundation DFG, via the Collaborative Research Center
SFB1491: Cosmic Interacting Matters - from Source to Signal.

\bibliography{papers.bib}

\begin{thebibliography}{}

\bibitem[{Abbott} et~al.,
  2017]{Abbott+2017_GW170817_MultiMessengerObservations}
{Abbott}, B.~P., {Abbott}, R., {Abbott}, T.~D., {Acernese}, F., {Ackley}, K.,
  {Adams}, C., {Adams}, T., {Addesso}, P., {Adhikari}, R.~X., and {Adya}, V.~B.
  (2017).
\newblock {Multi-messenger Observations of a Binary Neutron Star Merger}.
\newblock {\em \apj}, 848(2):L12.

\bibitem[{Alarcon} et~al.,
  2023]{Alarcon+2023_CMOS_QHY600_QHY411_characterization}
{Alarcon}, M.~R., {Licandro}, J., {Serra-Ricart}, M., {Joven}, E., {Gaitan},
  V., and {de Sousa}, R. (2023).
\newblock {Scientific CMOS sensors in Astronomy: QHY600 and QHY411}.
\newblock {\em arXiv e-prints}, page arXiv:2302.03700.

\bibitem[{Bellm} et~al., 2019]{Bellm+2019_ZTF_Overview}
{Bellm}, E.~C., {Kulkarni}, S.~R., {Graham}, M.~J., {Dekany}, R., {Smith},
  R.~M., {Riddle}, R., {Masci}, F.~J., {Helou}, G., {Prince}, T.~A., and
  {Adams}, S.~M. (2019).
\newblock {The Zwicky Transient Facility: System Overview, Performance, and
  First Results}.
\newblock {\em \pasp}, 131(995):018002.

\bibitem[{Ben-Ami}, 2023]{Ben-Ami+2023PASP_LAST_Science}
{Ben-Ami}, S. e.~a. (2023).
\newblock {The Large Array Survey Telescope -- Science goals}.
\newblock {\em submitted}.

\bibitem[{Bertin} and {Arnouts}, 1996]{Bertin+1996_SExtractor}
{Bertin}, E. and {Arnouts}, S. (1996).
\newblock {SExtractor: Software for source extraction.}
\newblock {\em \aaps}, 117:393--404.

\bibitem[{Bloemen} et~al., 2015]{Bloemen+Groot+2015ASPC_BlackGEM}
{Bloemen}, S., {Groot}, P., {Nelemans}, G., and {Klein-Wolt}, M. (2015).
\newblock {The BlackGEM Array: Searching for Gravitational Wave Source
  Counterparts to Study Ultra-Compact Binaries}.
\newblock In {Rucinski}, S.~M., {Torres}, G., and {Zejda}, M., editors, {\em
  Living Together: Planets, Host Stars and Binaries}, volume 496 of {\em
  Astronomical Society of the Pacific Conference Series}, page 254.

\bibitem[{Chambers} et~al., 2016]{Chambers+2016_PS1_Surveys}
{Chambers}, K.~C., {Magnier}, E.~A., {Metcalfe}, N., {Flewelling}, H.~A.,
  {Huber}, M.~E., {Waters}, C.~Z., {Denneau}, L., {Draper}, P.~W., {Farrow},
  D., {Finkbeiner}, D.~P., {Holmberg}, C., {Koppenhoefer}, J., {Price}, P.~A.,
  and {Rest}, A., e.~a. (2016).
\newblock {The Pan-STARRS1 Surveys}.
\newblock {\em arXiv e-prints}, page arXiv:1612.05560.

\bibitem[{Corbett} et~al., 2020]{Corbett+2020_SatellitesGlints}
{Corbett}, H., {Law}, N.~M., {Soto}, A.~V., {Howard}, W.~S., {Glazier}, A.,
  {Gonzalez}, R., {Ratzloff}, J.~K., {Galliher}, N., {Fors}, O., and {Quimby},
  R. (2020).
\newblock {Orbital Foregrounds for Ultra-short Duration Transients}.
\newblock {\em \apjl}, 903(2):L27.

\bibitem[{Drout} et~al., 2014]{Drout+2014_Rapidly_Evolving}
{Drout}, M.~R., {Chornock}, R., {Soderberg}, A.~M., {Sand ers}, N.~E.,
  {McKinnon}, R., {Rest}, A., {Foley}, R.~J., {Milisavljevic}, D., {Margutti},
  R., and {Berger}, E. (2014).
\newblock {Rapidly Evolving and Luminous Transients from Pan-STARRS1}.
\newblock {\em \apj}, 794(1):23.

\bibitem[{Gaia Collaboration} et~al., 2021]{GAIA+2021_GAIAEDR3_Summary_Content}
{Gaia Collaboration}, {Brown}, A.~G.~A., {Vallenari}, A., {Prusti}, T., {de
  Bruijne}, J.~H.~J., {Babusiaux}, C., {Biermann}, M., and {Creevey}, e.~a.
  (2021).
\newblock {Gaia Early Data Release 3. Summary of the contents and survey
  properties}.
\newblock {\em \aap}, 649:A1.

\bibitem[{Gaia Collaboration} et~al., 2016]{GAIA+2016_GAIA_mission}
{Gaia Collaboration}, {Prusti}, T., {de Bruijne}, J.~H.~J., {Brown}, A.~G.~A.,
  {Vallenari}, A., {Babusiaux}, C., {Bailer-Jones}, C.~A.~L., {Bastian}, U.,
  {Biermann}, M., {Evans}, D.~W., {Eyer}, L., {Jansen}, F., {Jordi}, C.,
  {Klioner}, S.~A., {Lammers}, U., {Lindegren}, L., {Luri}, X., {Mignard}, F.,
  {Milligan}, D.~J., and {Panem}, C. e.~a. (2016).
\newblock {The Gaia mission}.
\newblock {\em \aap}, 595:A1.

\bibitem[{Heinze} et~al., 2018]{Heinze+2018_ATLAS_VarStars}
{Heinze}, A.~N., {Tonry}, J.~L., {Denneau}, L., {Flewelling}, H., {Stalder},
  B., {Rest}, A., {Smith}, K.~W., {Smartt}, S.~J., and {Weiland}, H. (2018).
\newblock {A First Catalog of Variable Stars Measured by the Asteroid
  Terrestrial-impact Last Alert System (ATLAS)}.
\newblock {\em \aj}, 156(5):241.

\bibitem[{Ho} et~al., 2021]{Ho+2021_ZTF_RapidlyEvolvingTransients}
{Ho}, A. Y.~Q., {Perley}, D.~A., {Gal-Yam}, A., {Lunnan}, R., {Sollerman}, J.,
  {Schulze}, S., {Das}, K.~K., {Dobie}, D., {Yao}, Y., {Fremling}, C., {Adams},
  S., {Anand}, S., {Andreoni}, I., {Bellm}, E.~C., {Bruch}, R.~J., {Burdge},
  K.~B., {Castro-Tirado}, A.~J., {Dahiwale}, A., {De}, K., {Dekany}, R.,
  {Drake}, A.~J., {Duev}, D.~A., {Graham}, M.~J., {Helou}, G., {Kaplan}, D.~L.,
  {Karambelkar}, V., {Kasliwal}, M.~M., {Kool}, E.~C., {Kulkarni}, S.~R.,
  {Mahabal}, A.~A., {Medford}, M.~S., {Miller}, A.~A., {Nordin}, J., {Ofek},
  E., {Petitpas}, G., {Riddle}, R., {Sharma}, Y., {Smith}, R., {Stewart},
  A.~J., {Taggart}, K., {Tartaglia}, L., {Tzanidakis}, A., and {Winters}, J.~M.
  (2021).
\newblock {The Photometric and Spectroscopic Evolution of Rapidly Evolving
  Extragalactic Transients in ZTF}.
\newblock {\em arXiv e-prints}, page arXiv:2105.08811.

\bibitem[{Icecube Collaboration} et~al.,
  2017]{Icecube+2017A&A_Icecube_NeutrinoMultipletsSearch}
{Icecube Collaboration}, {Aartsen}, M.~G., {Ackermann}, M., {Adams}, J.,
  {Aguilar}, J.~A., {Ahlers}, M., {Ahrens}, M., {Al Samarai}, I., and
  {Altmann}, e.~a. (2017).
\newblock {Multiwavelength follow-up of a rare IceCube neutrino multiplet}.
\newblock {\em \aap}, 607:A115.

\bibitem[{Ivezi{\'c}} et~al., 2019]{Ivezic+2019_LSST_Survey}
{Ivezi{\'c}}, {\v{Z}}., {Kahn}, S.~M., {Tyson}, J.~A., {Abel}, B., {Acosta},
  E., {Allsman}, R., {Alonso}, D., {AlSayyad}, Y., {Anderson}, S.~F., {Andrew},
  J., {Angel}, J. R.~P., {Angeli}, G.~Z., and {Ansari}, Reza, e.~a. (2019).
\newblock {LSST: From Science Drivers to Reference Design and Anticipated Data
  Products}.
\newblock {\em \apj}, 873(2):111.

\bibitem[{Kochanek} et~al., 2017]{Kochanek+2017_ASASSN_VarStars}
{Kochanek}, C.~S., {Shappee}, B.~J., {Stanek}, K.~Z., {Holoien}, T.~W.~S.,
  {Thompson}, T.~A., {Prieto}, J.~L., {Dong}, S., {Shields}, J.~V., {Will}, D.,
  {Britt}, C., {Perzanowski}, D., and {Pojma{\'n}ski}, G. (2017).
\newblock {The All-Sky Automated Survey for Supernovae (ASAS-SN) Light Curve
  Server v1.0}.
\newblock {\em \pasp}, 129(980):104502.

\bibitem[{Kulkarni}, 2020]{Kulkarni2020arXiv_FollowupOpticalTransients}
{Kulkarni}, S.~R. (2020).
\newblock {Towards An Integrated Optical Transient Utility}.
\newblock {\em arXiv e-prints}, page arXiv:2004.03511.

\bibitem[{K{\"u}sters} et~al., 2020]{Kusters+2020SPIE_LightSorce_Calibration}
{K{\"u}sters}, D., {Bastian-Querner}, B., {Aldering}, G., {Blot}, S., {Boone},
  K., {Copin}, Y., {Hebecker}, D., {Karg}, T., {Kowalski}, M., {Lombardo}, S.,
  {Nordin}, J., and {Rubin}, D. (2020).
\newblock {SCALA upgrade: development of a light source for sub-percent
  calibration uncertainties}.
\newblock In {\em Society of Photo-Optical Instrumentation Engineers (SPIE)
  Conference Series}, volume 11447 of {\em Society of Photo-Optical
  Instrumentation Engineers (SPIE) Conference Series}, page 1144771.

\bibitem[{K{\"u}sters} et~al., 2022]{Kusters+2022SPIE_LightSourceCalibration}
{K{\"u}sters}, D., {Bastian-Querner}, B., {Aldering}, G., {Karg}, T.,
  {Kossatz}, M., {Kowalski}, M., and {Lombardo}, S. (2022).
\newblock {SCALA update: deci-percent laboratory spectro-radiometric NIST
  calibration transfer to new flux reference sensors}.
\newblock In {Evans}, C.~J., {Bryant}, J.~J., and {Motohara}, K., editors, {\em
  Ground-based and Airborne Instrumentation for Astronomy IX}, volume 12184 of
  {\em Society of Photo-Optical Instrumentation Engineers (SPIE) Conference
  Series}, page 121847V.

\bibitem[{Law} et~al., 2022]{Law+2022PASP_ArgusArray}
{Law}, N.~M., {Corbett}, H., {Galliher}, N.~W., {Gonzalez}, R., {Vasquez}, A.,
  {Walters}, G., {Machia}, L., {Ratzloff}, J., {Ackley}, K., {Bizon}, C.,
  {Clemens}, C., {Cox}, S., {Eikenberry}, S., {Howard}, W.~S., {Glazier}, A.,
  {Mann}, A.~W., {Quimby}, R., {Reichart}, D., and {Trilling}, D. (2022).
\newblock {Low-cost Access to the Deep, High-cadence Sky: the Argus Optical
  Array}.
\newblock {\em \pasp}, 134(1033):035003.

\bibitem[{LSST Science Collaboration} et~al.,
  2009]{Abell+2009_LSST_ScienceBook_Ver2}
{LSST Science Collaboration}, {Abell}, P.~A., {Allison}, J., {Anderson}, S.~F.,
  {Andrew}, J.~R., {Angel}, J. R.~P., {Armus}, L., {Arnett}, D., {Asztalos},
  S.~J., {Axelrod}, T.~S., {Bailey}, S., {Ballantyne}, D.~R., {Bankert}, J.~R.,
  {Barkhouse}, W.~A., {Barr}, J.~D., {Barrientos}, L.~F., {Barth}, A.~J.,
  {Bartlett}, J.~G., {Becker}, A.~C., {Becla}, J., {Beers}, T.~C., {Bernstein},
  J.~P., {Biswas}, R., {Blanton}, M.~R., {Bloom}, J.~S., {Bochanski}, J.~J.,
  {Boeshaar}, P., {Borne}, K.~D., {Bradac}, M., {Brandt}, W.~N., {Bridge},
  C.~R., {Brown}, M.~E., {Brunner}, R.~J., {Bullock}, J.~S., {Burgasser},
  A.~J., {Burge}, J.~H., {Burke}, D.~L., {Cargile}, P.~A., {Chand rasekharan},
  S., {Chartas}, G., {Chesley}, S.~R., {Chu}, Y.-H., {Cinabro}, D., {Claire},
  M.~W., {Claver}, C.~F., {Clowe}, D., {Connolly}, A.~J., {Cook}, K.~H.,
  {Cooke}, J., {Cooray}, A., {Covey}, K.~R., {Culliton}, C.~S., {de Jong}, R.,
  {de Vries}, W.~H., {Debattista}, V.~P., {Delgado}, F., {Dell'Antonio}, I.~P.,
  {Dhital}, S., {Di Stefano}, R., {Dickinson}, M., {Dilday}, B., {Djorgovski},
  S.~G., {Dobler}, G., {Donalek}, C., {Dubois-Felsmann}, G., {Durech}, J.,
  {Eliasdottir}, A., {Eracleous}, M., {Eyer}, L., {Falco}, E.~E., {Fan}, X.,
  {Fassnacht}, C.~D., {Ferguson}, H.~C., {Fernandez}, Y.~R., {Fields}, B.~D.,
  {Finkbeiner}, D., {Figueroa}, E.~E., {Fox}, D.~B., {Francke}, H., {Frank},
  J.~S., {Frieman}, J., {Fromenteau}, S., {Furqan}, M., {Galaz}, G., {Gal-Yam},
  A., {Garnavich}, P., {Gawiser}, E., {Geary}, J., {Gee}, P., {Gibson}, R.~R.,
  {Gilmore}, K., {Grace}, E.~A., {Green}, R.~F., {Gressler}, W.~J.,
  {Grillmair}, C.~J., {Habib}, S., {Haggerty}, J.~S., {Hamuy}, M., {Harris},
  A.~W., {Hawley}, S.~L., {Heavens}, A.~F., {Hebb}, L., {Henry}, T.~J.,
  {Hileman}, E., {Hilton}, E.~J., {Hoadley}, K., {Holberg}, J.~B., {Holman},
  M.~J., {Howell}, S.~B., {Infante}, L., {Ivezic}, Z., {Jacoby}, S.~H., {Jain},
  B., {R}, {Jedicke}, {Jee}, M.~J., {Garrett Jernigan}, J., {Jha}, S.~W.,
  {Johnston}, K.~V., {Jones}, R.~L., {Juric}, M., {Kaasalainen}, M.,
  {Styliani}, {Kafka}, {Kahn}, S.~M., {Kaib}, N.~A., {Kalirai}, J., {Kantor},
  J., {Kasliwal}, M.~M., {Keeton}, C.~R., {Kessler}, R., {Knezevic}, Z.,
  {Kowalski}, A., {Krabbendam}, V.~L., {Krughoff}, K.~S., {Kulkarni}, S.,
  {Kuhlman}, S., {Lacy}, M., {Lepine}, S., {Liang}, M., {Lien}, A., {Lira}, P.,
  {Long}, K.~S., {Lorenz}, S., {Lotz}, J.~M., {Lupton}, R.~H., {Lutz}, J.,
  {Macri}, L.~M., {Mahabal}, A.~A., {Mandelbaum}, R., {Marshall}, P., {May},
  M., {McGehee}, P.~M., {Meadows}, B.~T., {Meert}, A., {Milani}, A., {Miller},
  C.~J., {Miller}, M., {Mills}, D., {Minniti}, D., {Monet}, D., {Mukadam},
  A.~S., {Nakar}, E., {Neill}, D.~R., {Newman}, J.~A., {Nikolaev}, S.,
  {Nordby}, M., {O'Connor}, P., {Oguri}, M., {Oliver}, J., {Olivier}, S.~S.,
  {Olsen}, J.~K., {Olsen}, K., {Olszewski}, E.~W., {Oluseyi}, H., {Padilla},
  N.~D., {Parker}, A., {Pepper}, J., {Peterson}, J.~R., {Petry}, C., {Pinto},
  P.~A., {Pizagno}, J.~L., {Popescu}, B., {Prsa}, A., {Radcka}, V., {Raddick},
  M.~J., {Rasmussen}, A., {Rau}, A., {Rho}, J., {Rhoads}, J.~E., {Richards},
  G.~T., {Ridgway}, S.~T., {Robertson}, B.~E., {Roskar}, R., {Saha}, A.,
  {Sarajedini}, A., {Scannapieco}, E., {Schalk}, T., {Schindler}, R.,
  {Schmidt}, S., {Schmidt}, S., {Schneider}, D.~P., {Schumacher}, G.,
  {Scranton}, R., {Sebag}, J., {Seppala}, L.~G., {Shemmer}, O., {Simon}, J.~D.,
  {Sivertz}, M., {Smith}, H.~A., {Allyn Smith}, J., {Smith}, N., {Spitz},
  A.~H., {Stanford}, A., {Stassun}, K.~G., {Strader}, J., {Strauss}, M.~A.,
  {Stubbs}, C.~W., {Sweeney}, D.~W., {Szalay}, A., {Szkody}, P., {Takada}, M.,
  {Thorman}, P., {Trilling}, D.~E., {Trimble}, V., {Tyson}, A., {Van Berg}, R.,
  {Vand en Berk}, D., {VanderPlas}, J., {Verde}, L., {Vrsnak}, B., {Walkowicz},
  L.~M., {Wand elt}, B.~D., {Wang}, S., {Wang}, Y., {Warner}, M., {Wechsler},
  R.~H., {West}, A.~A., {Wiecha}, O., {Williams}, B.~F., {Willman}, B.,
  {Wittman}, D., {Wolff}, S.~C., {Wood-Vasey}, W.~M., {Wozniak}, P., {Young},
  P., {Zentner}, A., and {Zhan}, H. (2009).
\newblock {LSST Science Book, Version 2.0}.
\newblock {\em arXiv e-prints}, page arXiv:0912.0201.

\bibitem[{Nir} et~al., 2021a]{Nir+Ofek+2021_WFAST}
{Nir}, G., {Ofek}, E.~O., {Ben-Ami}, S., {Segev}, N., {Polishook}, D.,
  {Hershko}, O., {Diner}, O., {Manulis}, I., {Zackay}, B., {Gal-Yam}, A., and
  {Yaron}, O. (2021a).
\newblock {The Weizmann Fast Astronomical Survey Telescope (W-FAST): System
  Overview}.
\newblock {\em arXiv e-prints}, page arXiv:2105.03436.

\bibitem[{Nir} et~al., 2020]{Nir+2020_Satellites_Glints_FlaresLimit}
{Nir}, G., {Ofek}, E.~O., {Ben-Ami}, S., {Segev}, N., {Polishook}, D., and
  {Manulis}, I. (2020).
\newblock {A high-rate foreground of sub-second flares from geosynchronous
  satellites}.
\newblock {\em arXiv e-prints}, page arXiv:2011.03497.

\bibitem[{Nir} et~al., 2021b]{Nir+2021_RNASS_GN-z11-Flash_SatelliteGlint}
{Nir}, G., {Ofek}, E.~O., and {Gal-Yam}, A. (2021b).
\newblock {The GN-z11-Flash Event can be a Satellite Glint}.
\newblock {\em Research Notes of the American Astronomical Society}, 5(2):27.

\bibitem[{Ofek}, 2014]{Ofek2014_MAAT}
{Ofek}, E.~O. (2014).
\newblock {MATLAB package for astronomy and astrophysics}.

\bibitem[{Ofek}, 2019]{Ofek2019_Astrometry_Code}
{Ofek}, E.~O. (2019).
\newblock {A Code for Robust Astrometric Solution of Astronomical Images}.
\newblock {\em \pasp}, 131(999):054504.

\bibitem[{Ofek} et~al., 2021]{Ofek+2021_AT2018lqh_RapidlyEvolving}
{Ofek}, E.~O., {Adams}, S.~M., {Waxman}, E., {Sharon}, A., {Kushnir}, D.,
  {Horesh}, A., {Ho}, A., {Kasliwal}, M.~M., {Yaron}, O., {Gal-Yam}, A.,
  {Kulkarni}, S.~R., {Bellm}, E., {Masci}, F., {Shupe}, D., {Dekany}, R.,
  {Graham}, M., {Riddle}, R., {Duev}, D., {Andreoni}, I., {Mahabal}, A., and
  {Drake}, A. (2021).
\newblock {AT2018lqh and the nature of the emerging population of day-scale
  duration optical transients}.
\newblock {\em arXiv e-prints}, page arXiv:2109.10931.

\bibitem[{Ofek} and {Ben-Ami},
  2020]{Ofek+BenAmi2020_Grasp_SkySurvrys_CostEffectivness}
{Ofek}, E.~O. and {Ben-Ami}, S. (2020).
\newblock {Seeing-limited imaging sky surveys -- small vs. large telescopes}.
\newblock {\em arXiv e-prints}, page arXiv:2011.04674.

\bibitem[{Ofek}, 2023]{Ofek+2023ASP_LAST_PipelineI}
{Ofek}, E.~O. e.~a. (2023).
\newblock {The Large Array Survey Telescope -- Pipeline. I. Basic image
  reduction and sub visits coaddition}.
\newblock {\em submitted.}

\bibitem[{Quimby} et~al., 2007]{Quimby+2007_SN2005ap_SLSN}
{Quimby}, R.~M., {Aldering}, G., {Wheeler}, J.~C., {H{\"o}flich}, P.,
  {Akerlof}, C.~W., and {Rykoff}, E.~S. (2007).
\newblock {SN 2005ap: A Most Brilliant Explosion}.
\newblock {\em \apjl}, 668(2):L99--L102.

\bibitem[{Quimby} et~al., 2011]{Quimby+2011_SLSN}
{Quimby}, R.~M., {Kulkarni}, S.~R., {Kasliwal}, M.~M., {Gal-Yam}, A., {Arcavi},
  I., {Sullivan}, M., {Nugent}, P., {Thomas}, R., {Howell}, D.~A., {Nakar}, E.,
  {Bildsten}, L., {Theissen}, C., {Law}, N.~M., {Dekany}, R., {Rahmer}, G.,
  {Hale}, D., {Smith}, R., {Ofek}, E.~O., {Zolkower}, J., {Velur}, V.,
  {Walters}, R., {Henning}, J., {Bui}, K., {McKenna}, D., {Poznanski}, D.,
  {Cenko}, S.~B., and {Levitan}, D. (2011).
\newblock {Hydrogen-poor superluminous stellar explosions}.
\newblock {\em \nat}, 474(7352):487--489.

\bibitem[{Rivkin} et~al., 2021]{Rivkin+2021PSJ_DART_MissionRequirments}
{Rivkin}, A.~S., {Chabot}, N.~L., {Stickle}, A.~M., {Thomas}, C.~A.,
  {Richardson}, D.~C., {Barnouin}, O., {Fahnestock}, E.~G., {Ernst}, C.~M.,
  {Cheng}, A.~F., {Chesley}, S., {Naidu}, S., {Statler}, T.~S., {Barbee}, B.,
  {Agrusa}, H., {Moskovitz}, N., {Terik Daly}, R., {Pravec}, P., {Scheirich},
  P., {Dotto}, E., {Della Corte}, V., {Michel}, P., {K{\"u}ppers}, M.,
  {Atchison}, J., and {Hirabayashi}, M. (2021).
\newblock {The Double Asteroid Redirection Test (DART): Planetary Defense
  Investigations and Requirements}.
\newblock {\em PSJ}, 2(5):173--196.

\bibitem[{Sagiv} et~al., 2014]{Sagiv+2014_ULTRASAT}
{Sagiv}, I., {Gal-Yam}, A., {Ofek}, E.~O., {Waxman}, E., {Aharonson}, O.,
  {Kulkarni}, S.~R., {Nakar}, E., {Maoz}, D., {Trakhtenbrot}, B., {Phinney},
  E.~S., {Topaz}, J., {Beichman}, C., {Murthy}, J., and {Worden}, S.~P. (2014).
\newblock {Science with a Wide-field UV Transient Explorer}.
\newblock {\em \aj}, 147(4):79.

\bibitem[{Soumagnac} and {Ofek}, 2018]{Soumagnac+Ofek2018_catsHTM}
{Soumagnac}, M.~T. and {Ofek}, E.~O. (2018).
\newblock {catsHTM: A Tool for Fast Accessing and Cross-matching Large
  Astronomical Catalogs}.
\newblock {\em \pasp}, 130(989):075002.

\bibitem[{Steeghs} et~al., 2022]{Steeghs+2022MNRAS_GOTO_TelescopeSurvey}
{Steeghs}, D., {Galloway}, D.~K., {Ackley}, K., {Dyer}, M.~J., {Lyman}, J.,
  {Ulaczyk}, K., {Cutter}, R., {Mong}, Y.~L., {Dhillon}, V., {O'Brien}, P.,
  {Ramsay}, G., {Poshyachinda}, S., {Kotak}, R., {Nuttall}, L.~K., {Pall{\'e}},
  E., {Breton}, R.~P., {Pollacco}, D., {Thrane}, E., {Aukkaravittayapun}, S.,
  {Awiphan}, S., {Burhanudin}, U., {Chote}, P., {Chrimes}, A., {Daw}, E.,
  {Duffy}, C., {Eyles-Ferris}, R., {Gompertz}, B., {Heikkil{\"a}}, T.,
  {Irawati}, P., {Kennedy}, M.~R., {Killestein}, T., {Kuncarayakti}, H.,
  {Levan}, A.~J., {Littlefair}, S., {Makrygianni}, L., {Marsh}, T.,
  {Mata-Sanchez}, D., {Mattila}, S., {Maund}, J., {McCormac}, J., {Mkrtichian},
  D., {Mullaney}, J., {Noysena}, K., {Patel}, M., {Rol}, E., {Sawangwit}, U.,
  {Stanway}, E.~R., {Starling}, R., {Str{\o}m}, P., {Tooke}, S., {West}, R.,
  {White}, D.~J., and {Wiersema}, K. (2022).
\newblock {The Gravitational-wave Optical Transient Observer (GOTO): prototype
  performance and prospects for transient science}.
\newblock {\em \mnras}, 511(2):2405--2422.

\bibitem[{Tatum}, 1978]{Tatum1978_PolarAlignMethods}
{Tatum}, J.~B. (1978).
\newblock {Some methods for aligning the polar axis of a telescope.}
\newblock {\em Journal of the British Astronomical Association}, 89:21--37.

\bibitem[{Tonry}, 2011]{Tonry2011_ATLAS_SurveyCapability}
{Tonry}, J.~L. (2011).
\newblock {An Early Warning System for Asteroid Impact}.
\newblock {\em \pasp}, 123(899):58.

\bibitem[{Zackay} and {Ofek}, 2017a]{Zackay+2017_CoadditionI}
{Zackay}, B. and {Ofek}, E.~O. (2017a).
\newblock {How to COAAD Images. I. Optimal Source Detection and Photometry of
  Point Sources Using Ensembles of Images}.
\newblock {\em \apj}, 836(2):187.

\bibitem[{Zackay} and {Ofek}, 2017b]{Zackay+2017_CoadditionII}
{Zackay}, B. and {Ofek}, E.~O. (2017b).
\newblock {How to COAAD Images. II. A Coaddition Image that is Optimal for Any
  Purpose in the Background-dominated Noise Limit}.
\newblock {\em \apj}, 836(2):188.

\bibitem[{Zackay} et~al., 2016]{Zackay+2016_ZOGY_ImageSubtraction}
{Zackay}, B., {Ofek}, E.~O., and {Gal-Yam}, A. (2016).
\newblock {Proper Image Subtraction{\textemdash}Optimal Transient Detection,
  Photometry, and Hypothesis Testing}.
\newblock {\em \apj}, 830(1):27.

\bibitem[{Zhu} and {Dong},
  2021]{Zhu+Dong2021ARA&A_Exoplanets_Statistics_Implications}
{Zhu}, W. and {Dong}, S. (2021).
\newblock {Exoplanet Statistics and Theoretical Implications}.
\newblock {\em \araa}, 59.

\end{thebibliography}
\bibliographystyle{apalike}

\end{document}